\documentstyle [12pt,epsf] {article}
\def\vec#1{\mathchoice{\mbox{\boldmath$\mathrm\displaystyle#1$}}
{\mbox{\boldmath$\mathrm\textstyle#1$}}
{\mbox{\boldmath$\mathrm\scriptstyle#1$}}
{\mbox{\boldmath$\mathrm\scriptscriptstyle#1$}}}
\newcommand{\bm}[1]{\mbox{\boldmath$#1$}}  
\renewcommand{\vec}{\bm}
\newcommand{\be}{\begin{equation}}
\newcommand{\ee}{\end{equation}}
\newcommand{\bear}{\begin{eqnarray}}
\newcommand{\ear}{\end{eqnarray}}
\newcommand{\bea}{\begin{eqnarray*}}
\newcommand{\ea}{\end{eqnarray*}}
\newcommand{\lsim}{\mathrel{\vcenter
    {\hbox{$<$}\nointerlineskip\hbox{$\sim$}}}}
\newcommand{\gsim}{\mathrel{\vcenter
    {\hbox{$>$}\nointerlineskip\hbox{$\sim$}}}}

\newcommand{\Eins}{1\!\rm l}

\newcommand{\T}{\textstyle}

\newcommand{\simgt}{\hbox{ \raise3pt\hbox to 0pt{$>$}
    \raise-3pt\hbox{$\sim$} }}
\newcommand{\simsm}{\hbox{ \raise3pt\hbox to 0pt{$<$}
    \raise-3pt\hbox{$\sim$} }}

\begin{document}
\begin{flushright}
HD--THEP--98--33
\end{flushright}
\vspace{3cm}

\begin{center}
{\LARGE{ Differential cross sections
for high energy elastic
hadron-hadron scattering in nonperturbative QCD}\footnote{Supported by German 
Bundesministerium f\"ur Bildung und Forschung (BMBF),\\
\hphantom{dfd} Contract Nr. 05 7HD 91 P(0) 
and
by Deutsche Forschungsgemeinschaft
under \\ 
\hphantom{dfd} grant no. GKR 216/1-98}}
\vspace{2cm}

{\sc E. R. Berger\footnote{Email:
E.Berger@thphys.uni-heidelberg.de}},
{\sc O. Nachtmann\footnote{Email: O.Nachtman@thphys.uni-heidelberg.de}}
\\
\vspace{0.5cm}
{\em Institut f{\"u}r Theoretische Physik} \\
{\em Universit\"at Heidelberg} \\
{\em Philosophenweg 16} \\
{\em 69120 Heidelberg, Germany}

\end{center}
\vspace{1cm}

\thispagestyle{empty}

\begin{abstract}

Total and
differential cross sections for high energy and small
momentum transfer elastic hadron-hadron scattering are
studied in QCD using a functional integral approach.
The hadronic amplitudes are governed by  vacuum expectation
values of lightlike Wegner-Wilson loops, for which a matrix
cumulant expansion is derived. The cumulants are evaluated
within the framework of the Minkowskian version of the model
of the stochastic vacuum. Using the
second cumulant, we calculate elastic differential cross
sections for hadron-hadron scattering. The
agreement with experimental data is  good.

\end{abstract}

\newpage

\section{Introduction}

In this article we will discuss
elastic scattering of hadrons at high centre of
mass energy  $\sqrt s$ ($\sqrt s
\gsim
20$ GeV) and low momentum transfer squared $t$(say
$|t|{\mathrel{\vcenter
    {\hbox{$<$}\nointerlineskip\hbox{$\sim$}}}}
    \rm O (1\,{\rm GeV}^2)$).
Because of the small momentum transfer, such reactions are
governed by soft, nonperturbative interactions. Experimental
data show a rise for the total cross sections of all
hadronic reactions \cite{particle} with increasing centre of mass (c.m.)
energy, starting at about $\sqrt{\rm s} =  10 \, \rm GeV$.
Donnachie and Landshoff (DL) showed \cite{dola1} that this rise can
be described phenomenologically in terms of Regge theory \cite{regge}
by pomeron exchange.
The DL pomeron couples like a $C=+1$  ``photon'' to single quarks
in the hadrons. The transition from the quark to the hadron
level  leads then to the additive quark rule \cite{addrule}.
Donnachie and Landshoff
fitted the rise of all hadronic cross sections with one
small power of s, indicating that there is a universal mechanism
which governs this kind of reactions. There is also a lot of data
available for elastic differential cross sections  at different c.m.
energies, mainly for pp and p$\bar{\rm p}$ scattering
\cite{ppdat1,ppdat2,ppdat3}, but
also for $\pi \rm p$ and Kp scattering \cite{mbdat}. Surely the mechanism
which governs the elastic amplitude in the forward direction should
also control the elastic differential cross section (d$\sigma/dt$)
for sufficiently small $|t|$. Indeed, pomeron exchange is also able to
describe $d\sigma /dt$ \cite{dola2}. There are many other
proposals and methods to describe such hadronic reactions,
from perturbative field theoretic calculations \cite{perturb},
topological expansions and strings \cite{strings},
valons \cite{valons}, the work of Cheng and Wu on high
energy behaviour in field theories based on perturbation theory
\cite{cheng}, to
``geometrical'' models, which invoke global phenomenological
properties of hadrons like their ``blackness'' \cite{geometry}.
For a review of ``pomeron physics'' we refer to \cite{12a}.

A new effort
towards a microscopic description of high energy soft hadronic
reactions was made in \cite{lana}.
In an abelian gluon model the pomeron properties were related to
nonperturbative properties of the vacuum like the gluon condensate
\cite{sum} and a ``vacuum correlation length'' $a$ \cite{14a}.
In \cite{na91} these ideas were generalised
for QCD. It was shown there that the amplitude
for qq-scattering at high energies
is governed by the correlation
function of two lightlike Wegner-Wilson lines.
Using this formalism a description of hadron-hadron scattering
was developed in \cite{15a,dfk} where the hadronic amplitudes are
calculated from correlation functions of lightlike
Wegner-Wilson loops.
These correlation functions are evaluated  in the model of the
stochastic vacuum \cite{msv}, applied in Minkowski space
after an analytic continuation from Euklidean space.
There are by now many other applications of this technique,
for example in the
description of  exclusive vector meson production \cite{pir}.
Related techniques have been used in \cite{heb}
for dealing with 
hard diffractive
processes in deep inelastic lepton-nucleon scattering observed
at HERA  \cite{251}. \\[0.005cm]

The goal of our paper is, to use and further develop the description
of high energy diffractive hadron scattering given
in \cite{na91, dfk, na96, 15b}.
In Sect. 2 we collect the formulae for the hadronic scattering
amplitudes as derived there.
We begin Sect. 3 with a summary of the model of the stochastic
vacuum (MSV) in its Minkowski version. In the second part of Sect. 3
we calculate the correlation function of two Wegner-Wilson loops, the
main ingredient of the meson-meson scattering amplitude, using a
matrix cumulant expansion and the MSV.  A question of interest is
whether or not the constituents of the baryons prefer quark-diquark like
configurations where two quarks are close to each other on a scale
given by the proton radius.  In \cite{odderon,dnr} strong arguments
for the quark-diquark picture were given where baryons do
act in a first approximation as colour dipoles in the same way as
mesons.  In  Sect. 4 we present our results for the
$pp$ and $p\bar p$ elastic differential cross section
$d \sigma / dt$ treating the baryons as such colour dipoles.
In Sect. 5 we discuss meson-meson and meson-baryon scattering
considering baryons again as colour dipoles.  Treating
meson-baryon and baryon-baryon scattering for general
three-quark baryon configurations
along these lines is more complicated and we defer this to
another publication. Our conclusions are drawn in  Sect. 6.

\section{The hadronic amplitudes}
\label{The hadronic amplitudes}

\begin{figure}[htb]

\vspace*{-0.5cm}

\hspace{1cm}
\epsfysize=7cm
\centerline{\epsffile{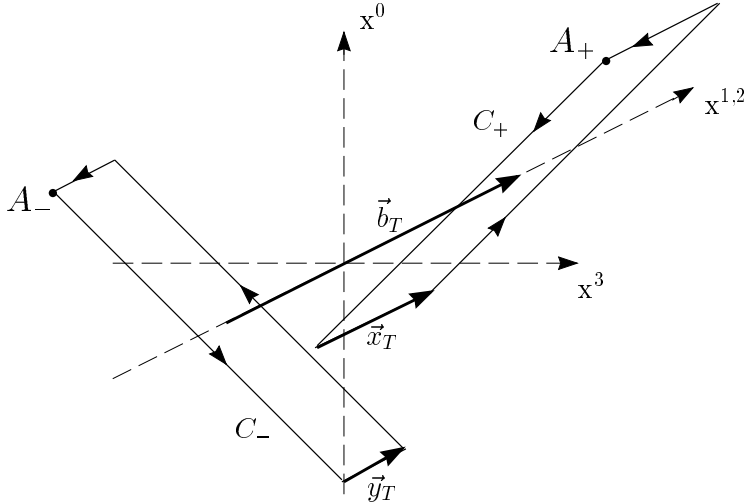}}

\vspace*{0cm}

\caption[a]{   \em The light-like Wegner-Wilson cut
loops in Minkowski space time,
$C_{\pm}$, consisting of two light like lines in the hyperplanes
$x_{\mp}=0$ and connecting pieces at infinity.
The loops are cut open at one corner, $A_-$ and $A_+$, respectively.
In transverse space
the centres of the loops are at $\pm \vec{b}_T/2$,
the vectors from the antiquark to the quark lines are $\vec{x}_T$
and   $\vec{y}_T$ respectively.}
\label{mbild}
\end{figure}

Consider elastic scattering of two hadrons $h_1,h_2$ in the c.m. system
at high energies and small momentum transfer
\begin{equation}
  h_1( P_1)+h_2( P_2) \rightarrow h_1( P_3)+h_2( P_4).
\label{reactions}
\end{equation}
%
%
Now look at this reaction with a microscope. We have to choose an
appropriate resolution in order to extract the essential features of
the reaction, but not resolve unimportant details. In \cite{na91} this
resolution was estimated, based on the uncertainty relation, with the
following conclusions. Over a time interval $t_0 \approx 2 \, {\rm fm}$
around the ``nominal'' interaction time (i) the parton state of the
hadrons does not change qualitatively, i.e. parton annihilation and
production processes can be neglected, (ii) partons travel in essence
on straight lightlike world lines and (iii) the partons undergo
``soft'' elastic scattering governed by non-perturbative gluon
dynamics. Using this approach hadronic amplitudes for high energies
were derived in \cite{dfk,na96}. The result for
meson-meson scattering where mesons are represented as
$q\bar{q}$ wave packets is
\begin{eqnarray}
  &&S_{fi} = \delta_{fi} + i(2\pi)^4 \delta (P_3+P_4-P_1-P_2) T_{fi},
  \nonumber\\[.05cm]
  &&T_{fi} = (-2is) \int d^2b_T {\exp}(i{\bm q}_T{\bm b}_T)
  \int d^2x_T d^2y_T \, w_{3,1}^M({\bm {x}}_T) \, w_{4,2}^M({\bm {y}}_T)
  \nonumber\\
  &&\hphantom{T_{fi} =}
  \Big\langle W_+^M(\frac{1}{2} {\bm b}_T,{\vec x}_T)
  W_-^M(-\frac{1}{2} {\vec b}_T,{\vec y}_T)-1 \Big\rangle_G .
  \label{mamp}
\end{eqnarray}
%
%
Here the assumption is made that the $q$ and $\bar{q}$
share the longitudinal momentum of the meson
roughly in equal proportions.
The interpretation of (\ref{mamp}) and the symbols
occurring there is as follows.
The scattering amplitude is obtained by first considering the
scattering of quarks and antiquarks on a fixed gluon potential
and then summing over all
gluon potentials by path integration, indicated with the brackets
$\langle \; \rangle_G$. Travelling through a gluon potential the
quarks and antiquarks pick up non-abelian phase factors.
To ensure  gauge invariance the phase factors for
$q$ and $\bar q$ from the  same meson
are joined at large positive and negative
times, yielding lightlike Wegner-Wilson loops $W_{\pm}$
which are defined as
\be
\label{wloop}
  W_{\pm}^M \equiv \frac{1}{3} \, {\rm{tr}} \, V(C_{\pm}),
 \ee
%
%
 \be 
 V(C_{\pm}) =  {\rm P}\ {\rm exp} [ -ig\int_{\it{C_{\pm}}}
 dx^{\mu} \, G_{\mu}^{a}(x) \frac{\lambda^{a}}{2} ].
\ee
%
%
Here $V(C_{\pm})$ are non-abelian phase factors (connectors)
along cut loops $ C_{\pm}$ as sketched in Fig. \ref{mbild}.
The trace in (\ref{wloop}) corresponds in the usual way to the closure
of the loop.
The transverse separation
between the centres of the loops is given by $\vec{b}_T$. The
vectors $\vec{x}_T$ and $\vec{y}_T$ give the extensions and
orientations of the loops in transverse space.
The path integration correlates
these loops and so causes the interaction.
The resulting loop-loop
correlation function  has to be integrated over all extensions and
orientations of the loops in transverse space  with a measure given by
the meson's overlap functions $w_{3,1}^M$ and
$w_{4,2}^M$ for which one has
to make a suitable ansatz.
In order to obtain the hadronic amplitude  a
Fourier transform over the impact parameter
$\vec{b}_T$ has to be done finally. In the following we call
(\ref{mamp}) the meson-meson amplitude.
As discussed in the introduction, we  can
use it  also to describe
meson-baryon and baryon-baryon scattering treating
baryons as colour dipoles in the quark-diquark picture.

\section{Evaluation of the meson-meson  scattering amplitude}

In this section we will use the model of the
stochastic vacuum (MSV) to perform the functional integral
in (\ref{mamp}) in an approximate way.

First we give a short summary of the relevant properties
of the MSV. For a detailed
discussion see \cite{dfk,msv}. The most important ingredient
is a special ansatz for the correlation function of two
parallel transported gluon field strength tensors, shifted to a
common reference point $o$ along the curves $C_{x_1}$ and $C_{x_2}$.
\begin{equation}
  \Big\langle \frac{g^2}{4 \pi^2}
   \hat{G}_{\mu \nu}^a (o,x_1;{\it{C}}_{x_1})
  \hat{G}_{\rho\sigma}^b (o,x_2;{\it{C}}_{x_2})
  \Big\rangle \equiv
  \frac{1}{4} \delta^{ab} \, F_{\mu \nu \rho \sigma}
  (x_1,x_2,o;C_{x_1},C_{x_2}).
  \label{correlator}
\end{equation}
%
%
Here the right-hand side of (\ref{correlator}) depends on
$x_1,x_2$ and $C_{x_1},C_{x_2}$. The reference point $o$ can be freely
shifted on the curve $C_{12} \equiv C_{x_1} + \bar{C}_{x_2}$
where $\bar{C}_{x_2}$ is the oppositely
oriented curve $C_{x_2}$. The
correlation function is proportional to $\delta^{ab}$ due to colour
conservation. Now the MSV makes the assumption, that
$ F_{\mu \nu \rho \sigma}$ is independent of the connecting curve
$C_{12}$. Then Poincar$\acute{{\rm{e}}}$ and parity invariance require the
correlator to be of the following form $(z=x_1-x_2)$
\begin{eqnarray}
   &&F_{\mu \nu \rho \sigma}(z)=
   \frac{1}{24} G_2
   \{ (g_{\mu \rho} g_{\nu \sigma}-
   g_{\mu \sigma} g_{\nu \rho}) [\kappa \, D(z^2)
   +(1-\kappa)D_1(z^2)] \nonumber\\
   && \hphantom{F_{\mu \nu \rho \sigma}(z)= } +
       (z_{\sigma}z_{\nu} g_{\mu \rho} -
    z_{\rho}z_{\nu} g_{\mu \sigma} +
    z_{\rho}z_{\mu}g_{\nu \sigma} -
    z_{\sigma}z_{\mu}g_{\nu \rho} )
    (1 - \kappa) \frac{{\rm{d}} D_1(z^2)}{{\rm{d}}(z^2)}\} .
   \label{tracecorrelator}
\end{eqnarray}
%
%
Here $G_2$ is the gluon condensate, $D$ and $D_1$ are invariant functions
normalized to $D(0)=D_1(0)=1$. For spacelike separations they are
assumed to fall off rapidly on a length scale given by the
correlation length $a \simeq 0.3 \ \rm{fm}$.
The Fourier decomposition of the correlation
functions is given by
\begin{eqnarray}
   D(z^2)  &=&
   \int_{-\infty}^{\infty}
  \frac{d^4k}{(2\pi)^4} e^{-ikz}
  \tilde{D}(k^2) ,
  \nonumber\\
   D_1(z^2)  &=&
   \int_{-\infty}^{\infty}
  \frac{d^4k}{(2\pi)^4} e^{-ikz}
  \tilde{D}_1(k^2) .
\label{corfktmom1}
\end{eqnarray}
%
%
We follow the authors of \cite{dfk} and take as
ansatz for $\tilde{D}$ and $\tilde{D}_1$
\begin{eqnarray}
   \tilde{D}(k^2)  &=&  \frac{27(2\pi)^4}{(8a)^2}
  \frac{i k^2}{(k^2 - \lambda^{-2} + i\epsilon)^4},
  \nonumber\\
   \tilde{D}_1(k^2)  &=& \frac{2}{3}
  \frac{27(2\pi)^4}{(8a)^2}
  \frac{i}{(k^2 - \lambda^{-2} + i\epsilon)^3} .
\label{corfktmom}
\end{eqnarray}
%
%
The constant $\lambda$ is given by $ \lambda=8 \, a/ 3 \, \pi$
and $\kappa$ is a
parameter related to  the non-abelian character of the correlator
\cite{dfk,25a}.

The Euclidean version of the correlator (\ref{correlator}) has been
investigated in lattice QCD \cite{gitter}. The ansatz
(\ref{corfktmom1}), (\ref{corfktmom}) gives a good description
of the nonperturbative part of the correlator in comparison to
the data from the measurements in quenched QCD and from a fit
one finds the following ranges for the parameters $G_2,a,\kappa$
\cite{megg}:
\bear
\label{10a}
\kappa G_2a^4&=&0.39\ {\rm to}\ 0.41,\nonumber\\
\kappa&=&0.80\ {\rm to}\ 0.89,\nonumber\\
a&=&0.33\ {\rm to}\ 0.37\ {\rm fm}.
\ear
%
%
From the lattice data one obtains directly the dimensionless
quantities $\kappa$ and $\kappa G_2a^4$. Their uncertainty
is due to statistical errors of the lattice data and to
variations of the fit range chosen. The values for $a$ depend
also on $\Lambda_{\rm Lattice}$ which introduces some further
uncertainty.

It was shown in \cite{msv} that
$\kappa \not= 0$ is crucial for deriving confinement in the
framework of the MSV. As
was found for the total cross sections in \cite{dfk} and as we
will  find again here a value $\kappa \not= 0$
is also crucial  to
describe the experimental data on high energy scattering
for $d\sigma / dt$ in the framework of our model.

\subsection{A cumulant expansion
of the colour dipole  correlation function}

Here we calculate the colour dipole correlation
function $\langle W_+^M W_-^M \rangle_G $ introduced in (\ref{mamp}).
%
%
\begin{figure}[htb]

\vspace*{-0.5cm}

\hspace{1cm}
\epsfysize=7cm
\centerline{\epsffile{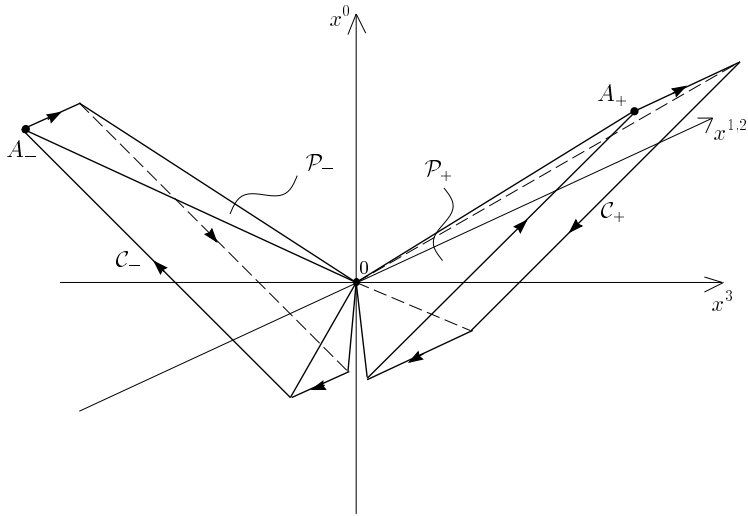}}

\vspace*{0cm}

\caption[a]{ \em The curves $C_+$ and $C_-$ along which the
connectors in $W_{\pm}$ are taken. The mantle of the pyramid
with apex at the origin of the coordinate system and
boundary $C_+(C_-)$ is $P_+(P_-)$, the basis surface $S_+(S_-)$.}
\label{pbild}
\end{figure}
The strategy is the following. We transform the line integrals
$W_+^M$ and  $W_-^M$ into a surface integral using the non-abelian
Stokes theorem \cite{stokes}. Following \cite{dfk} we
choose as surface the mantle of
the double pyramid $P = P_+ + P_-$ which has $C_+ + C_-$
as boundary and $o$ as apex (Fig. \ref{pbild}).

To give the details, let $P_\pm$ be the pyramid surfaces
excluding the base surfaces  $S_\pm$ and cut from o to
$A_\pm$. Then the boundaries of $P_\pm$ are
\bear
\label{10b}
\partial P_+&=&oA_++C_++A_+o,\nonumber\\
\partial P_-&=&oA_-+C_-+A_-o.
\ear
%
%
%
The methods developed for the non-abelian Stokes theorem as
explained in \cite{na96} allow us to write the line integrals
of (\ref{wloop}) as
\bear
\label{10c}
V(C_+)&=&V(A_+o)V(P_+)V(oA_+),\nonumber\\
V(C_-)&=&V(A_-o)V(P_-)V(oA_-).
\ear
%
%
Here $V(oA_\pm),\ V(A_\pm o)$ are connectors
along the straight lines from $A_\pm$ to $o$ and $o$ to
$A_\pm$, respectively. They satisfy
\bear
\label{10d}
&&V(oA_+)V(A_+o)=\Eins\nonumber\\
&&V(oA_-)V(A_-o)=\Eins
\ear
%
%
The matrices $V(P_\pm)$ in (\ref{10c}) are surface-ordered
exponentials of field strength tensors $\hat G$ parallel transported
to $o$:
\bear
\label{10e}
V(P_+)&=& {\rm P} \ \exp\left[-\frac{ig}{2}
\int_{P_+}d\sigma^{\mu\nu}(x)\hat  
G^a_{\mu\nu}(o,x;C_x)\frac{\lambda^a}{2}\right],\nonumber\\
V(P_-)&=& {\rm P} \ \exp\left[-\frac{ig}{2}
\int_{P_-}d\sigma^{\mu\nu}(x)\hat  
G^a_{\mu\nu}(o,x;C_x)\frac{\lambda^a}{2}\right].
\ear
%
%
Inserting (\ref{10c}) in (2-3), using (\ref{10d}) and
the cyclicity of the trace, we get
\begin{eqnarray}
  \label{tloopp}
  &&\Big\langle W_+^M(\frac{1}{2} \vec{b}_T,\vec{x}_T) \,
  W_-^M (-\frac{1}{2} \vec{b}_T,\vec{y}_T)\Big\rangle_G \equiv
  \Big\langle W^M_+W^M_-\Big\rangle_G
  \nonumber\\
  &&=\Big\langle\frac{1}{3}[{\rm tr} V(C_+)]\frac{1}{3}[{\rm tr}
  V(C_-]\Big\rangle_G\nonumber\\
  &&=\Big\langle\frac{1}{3}[{\rm tr} V(P_+)]\frac{1}{3}[{\rm tr}
  V(P_-]\Big\rangle_G.
\ear
%
%
The main idea is now to interpret
the product of the two traces (tr) over $3\times 3$ matrices
in (\ref{tloopp}) as one trace
$({\rm{Tr}}_2)$ acting in the $9$-dimensional
tensor product space carrying the product of two SU(3) quark
representations.
Using the definition of the matrix multiplication in the product
space giving e.g.
\begin{equation}\begin{array}{rcl}
  (\lambda^a \otimes 1) (\lambda^b \otimes 1) &=&
  \lambda^a \lambda^b \otimes 1 ,       \\
  (\lambda^a \otimes 1) (1 \otimes \lambda^b) &=&
  \lambda^a \otimes \lambda^b
  \label{matrixmulti}
\end{array}\end{equation}
%
%
and  of path ordering we get immediately
\begin{eqnarray}
  &&\Big\langle W_+^M \,   W_-^M \Big\rangle_G = 
  \frac{1}{9} \, {\rm{Tr}}_2
  \Big\langle \, {\rm P} \, {\rm{exp}} 
  [ -\frac{ig}{2} \int_{\it{P_+}}
  d\sigma^{\mu \nu}
  \, \hat{G}_{\mu \nu}^{a} 
  (\frac{\lambda^{a}}{2} \otimes 1) ]
  \nonumber\\
  &&\hphantom{\langle W_+^M \,  \; \; W_-^M 
    \rangle_G = \frac{1}{9} {\rm{Tr}}_2
    \langle \, }
  {\rm P} \, {\rm{exp}} [ -\frac{ig}{2} \int_{\it{P_-}}
  d\sigma^{\mu \nu}
  \, \hat{G}_{\mu \nu}^{a} (1 \otimes \frac{\lambda^{a}}{2}) ]
  \; \Big\rangle_G.
  \label{tloopf}
\end{eqnarray}
%
%
The two exponentials in (\ref{tloopf}) commute because the two matrix
structures in the exponents do. Introducing a total shifted
field strength tensor $\hat{G}_t$ as
\begin{equation}
  \hat{G}_{t,\mu \nu}(o,x;C_x)
  = \left\{
    \begin{array}{r@{\quad}l}
      \hat{G}^a_{\mu \nu}(o,x;C_x)
      (\frac{\lambda^a}{2} \otimes  1) & {\rm{for}}\quad x \,\epsilon \,
      P_+\\
      \hat{G}^a_{\mu \nu}(o,x;C_x)
      (1 \otimes \frac{\lambda^a}{2} ) & {\rm{for}}\quad x \, \epsilon \,
      P_-
    \end{array}
  \right.
  \label{gtot}
\end{equation}
%
%
we can rewrite the two exponentials in (\ref{tloopf})
as one exponential defined in the direct product space.
In this way we get from (\ref{tloopp}) a path-ordered integral
over the double  pyramid mantle $P=P_++P_-$:
\begin{equation}
  \Big\langle W_+^M \, W_-^M \Big\rangle_G =
  \frac{1}{9} {\rm{Tr}}_2 \Big\langle {\rm P}\ {\rm{exp}} [ -i\frac{g}{2}
  \int_{P}
  d\sigma (x)  \, \hat{G}_t (x) ]
  \Big\rangle_G .
  \label{sloop}
\end{equation}
%
%
Here and in the following
we supress the Lorentz indices if there is no confusion. Note that
the path orderings on $P_+$ and $P_-$ do not interfer
with each other. Thus the path-ordering on $P$ can for instance
be chosen such that all points of $P_+$ are ``later''
than all points of $P_-$.

For the expectation value of the
single surface ordered exponential (\ref{sloop}) we can
make a cumulant expansion \cite{vankamp,na96}.
In our case we use a matrix cumulant expansion
as explained in (2.41) of \cite{na96} (cf. also \cite{kala}):
\begin{equation}
  \Big\langle {\rm P}\, {\rm{exp}} [ -i\frac{g}{2}
  \int_{P}
  d \sigma (x)  \, \hat{G}_t (x) ] \Big\rangle_G  =
  {\rm{exp}} [\, \,  \sum_{n=1}^{\infty} \frac{1}{n !}
  (-i\frac{g}{2})^n
  \int d \sigma (x_1) \cdots d\sigma (x_n) \, K_n(x_1,..,x_n)] .
  \label{cumulantexpansion}
\end{equation}
%
%
Here the cumulants $K_n$ are functional integrals
over products of the
non-commuting matrices $\hat{G}_t$ of (\ref{gtot}). Thus
one has to be
careful with their ordering.
The cumulants up to $n=2$ are
\begin{eqnarray}
  K_1(x) &=& \big\langle \hat{G}_t(o,x;C_x) 
  \big\rangle_G, \nonumber\\
  K_2(x_1,x_2) &=& \big\langle {\rm P}
  [\hat{G}_t(o,x_1;C_{x_1})\hat{G}_t(o,x_2;C_{x_2})\, ]
    \big\rangle_G - \nonumber\\
    &&\frac{1}{2}\Big
    (\big\langle \hat{G}_t (o,x_1;C_{x_1}) \big\rangle_G
    \big\langle \hat{G}_t(o,x_2;C_{x_2}) \big\rangle_G +
    (1 \leftrightarrow 2)\, \Big) .
  \label{cumulants}
\end{eqnarray}
%
%
Note that the $\hat{G_t}$ have Lorentz indices and
are matrices in colour space as shown in (\ref{gtot}).
The functional  integral indicated by $\langle \; \rangle_G$
in (\ref{cumulants}) involves only the field strength
components $\hat{G}_{\mu \nu}^a$, thus also the cumulants
$K_1,\, K_2,\, ...$ still carry Lorentz and colour indices.
The fact that there is no colour direction
preferred in the vacuum requires $K_1$ to vanish. Neglecting
cumulants higher than $n=2$ we get for (\ref{tloopp})
\begin{eqnarray}
  &&\Big\langle W_+^M \,  W_-^M \Big\rangle_G =
  \frac{1}{9} {\rm{Tr}}_2 \, {\rm{exp}}(C_2(\vec x_T,\vec y_T,\vec b_T)),
  \nonumber\\
  && C_2 (\vec x_T,\vec y_T,\vec b_T)=
  -\frac{g^2}{8} \int_P d\sigma(x_1)
  \int_P d\sigma(x_2)\, \,
  \Big\langle {\rm P} (\hat{G}_t(o,x_1;C_{x_1})
  \hat{G}_t(o,x_2;C_{x_2}))
  \Big\rangle_G .
  \label{mcum}
\end{eqnarray}
%
%
where $C_2$ is a $9\times 9$ matrix, invariant under SU(3)
colour rotations.

\subsection{Calculation of the second cumulant term using the MSV}

Now we use the MSV  with the ansatz
(\ref{corfktmom1}),(\ref{corfktmom})
for the invariant functions
$D$ and $D_1$  to calculate
$C_2$ of (\ref{mcum}), which can be split into three contributions:
\begin{equation}
  C_2 = C_2^{P_+ P_+} + C_2^{P_- P_-} + C_2^{P_+ P_-}.
  \label{c2cont}
\end{equation}
%
%
In each of
$C_2^{P_+ P_+}$ and $C_2^{P_- P_-}$ both $x_1$
and $x_2$ move on the same
surface, $P_+$,  $P_-$, respectively,
and so we have to pay attention to the
surface ordering. In $C_2^{P_+ P_-}$ one point moves on $P_+$, the other
one on
$P_-$. In this case it follows from (\ref{gtot}) that the shifted
field strengths commute and the surface ordering
is irrelevant for $C_2^{P_+ P_-}$.

Now we show that
$C_2^{P_+ P_+}$ vanishes. To see this we follow the
argumentation of \cite{na96}
and transform the two surface integrals over the
pyramid mantle
$P_+$ in (\ref{mcum}) into surface integrals over $S_+$
and integrals over the volume $V_+$ enclosed
by $P_+$ and $S_+$.
The integrals over $S_+$  vanish  due to the Lorentz structure
of the surface elements $d \sigma^{\mu \nu }$
together with the ansatz (\ref{tracecorrelator}).
The integrals over $V_+$  are roughly speaking sums of integrals over
surfaces parallel to $S_+$ and so they vanish for the same reason.
In a similar way we see that $C_2^{P_- P_-}$ vanishes.

To calculate $C_2^{P_+ P_-}$ we use
the same method. In this case neither the
integrals over $S_{\pm}$ nor the ones over $V_{\pm}$ vanish.
The integrations in light-like directions can be done
analytically. Using then also the
ordinary Gauss theorem we find with
(\ref{correlator})-(\ref{corfktmom1}) that everything reduces to
line integrals over the vectors $\vec r_{xi},\ \vec r_{yi}\ (i=q,\bar q)$
running from the apex $o$ to the position of the quarks and
antiquarks in transverse space (Fig. \ref{rbild}):
\begin{eqnarray}
  &&C_2(\vec{x}_T,\vec{y}_T,\vec{b}_T) =
  \frac{\lambda^a}{2} \otimes \frac{\lambda^a}{2}
  (-i)\, \chi (\vec{x}_T,\vec{y}_T,\vec{b}_T),
  \nonumber\\
  &&\chi (\vec{x}_T,\vec{y}_T,\vec{b}_T) =
  \, \frac{ \, G_2 \, \pi^2}{24} \;
  \Big\{
    I( \vec{r}_{xq}, \vec{r}_{yq} ) +
    I( \vec{r}_{x\bar{q}}, \vec{r}_{y\bar{q}} ) -
    \nonumber\\
    &&\hphantom{\chi (\vec{x}_T,\vec{y}_T,\vec{b}_T) =
      i \, \frac{ \, G_2 \, \pi^2}{24} \;
      \Big\} }
    I( \vec{r}_{xq}, \vec{r}_{y\bar{q}} ) -
    I( \vec{r}_{x\bar{q}},\vec{r}_{yq} ) \; \; \Big\},
    \nonumber\\[0.01cm]
    && I(\vec{r}_{x}, \vec{r}_{y} ) =
    i\ \int_0^1\,d v_1 \int_0^1\,d v_2
    \; \int_{-\infty}^{\infty}
    \frac{d^2k_T}{(2\,\pi)^2}
    e^{{\T i \vec{k}_T(v_1\,\vec{r}_{y} - v_2\, \vec{r}_{x}) }}
    \nonumber\\
    && \hphantom{I(\vec{r}_{x}, \vec{r}_{y} ) =}
    \Big\{ \kappa \, \vec{r}_{y} \vec{r}_{x}
    \tilde{D}(-\vec{k}_T^2) +
    (1-\kappa)\, (\vec{k}_T \vec{r}_{y})
    (\vec{k}_T \vec{r}_{x})\;
    \tilde{D}_1'  (-\vec{k}_T^2) \Big\},
    \nonumber\\[0.02cm]
    && \tilde{D}_1'(k^2) = \frac{d}{dk^2} \tilde{D}_1(k^2).
    \label{mchi2}
  \end{eqnarray}
%
%
As we can see we finally need the correlator functions
$\tilde D,\tilde D_1$ of (\ref{tracecorrelator})
for space-like momenta only. This means that the result (\ref{mchi2})
involves the correlation functions $D(z^2)$ and $D_1(z^2)$
only for spacelike $z$, where they are as in Euclidean space time.
With (\ref{corfktmom}) for the functions $\tilde D, \, \tilde D_1$
we get
\begin{eqnarray}
  && I(\vec{r}_{x}, \vec{r}_{y} ) =
  \bigg\{ \kappa \frac{\pi}{2}\, \lambda^2 \;
  (\vec{r}_{y}\vec{r}_{x}) \; \int_{0}^{1} dv
  \big(\, (\frac{|v\vec{r}_{y}-\vec{r}_{x}|}{\lambda})^2
  K_2(\frac{|v\vec{r}_{y}-\vec{r}_{x}|}{\lambda}) +
  \nonumber\\
  &&\hphantom{
    I(\vec{r}_{x}, \vec{r}_{y} ) =
    \bigg\{ \kappa \frac{\pi}{2}\, \lambda^2 \;
    (\vec{r}_{y}\vec{r}_{x}) \; \int_{0}^{1} dv \big( }
  (\frac{|\vec{r}_{y}-v\vec{r}_{x}|}{\lambda})^2
  K_2(\frac{|\vec{r}_{}-v\vec{r}_{x}|}{\lambda})
  \big)
  \nonumber\\
  && \hphantom{
    I(\vec{r}_{x}, \vec{r}_{y} ) = }
  + (1-\kappa) \, \pi \, \lambda^4 \; \Big(
  (\frac{|\vec{r}_{y}-\vec{r}_{x}|}{\lambda})^3
  K_3(\frac{|\vec{r}_{y}-\vec{r}_{x}|}{\lambda})
  \; \; \bigg\}
  \label{mchi}
\end{eqnarray}
%
%
where $K_{2,3}$ are the modified Bessel functions. Note that
$\chi$ is a real function.
\\[0.001cm]
\begin{figure}[htb]

\vspace*{-0.5cm}

\hspace{1cm}
\epsfysize=6cm
\centerline{\epsffile{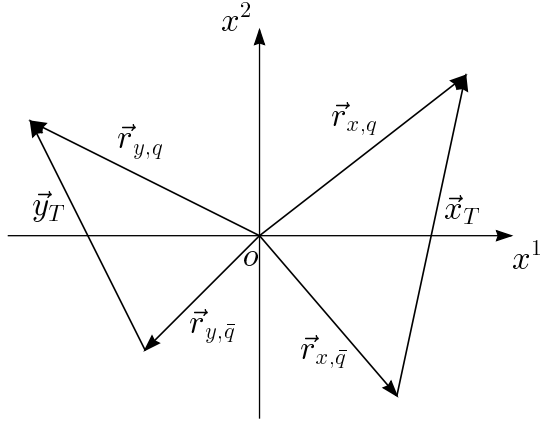}}

\vspace*{0cm}

\caption[a]{ \em The projection of the pyramid surfaces
$P_{\pm} + S_{\pm}$ into transverse space. The
vectors $\vec{r}_{x,q(\bar{q})}$ and
$\vec{r}_{y,q(\bar{q})}$ point from the origin o along the
transverse projections of the pyramid mantles
$P_{\pm}$ to the projections of the quark and antiquark
lines of $C_{\pm}$. $\vec{x}_T$ and $\vec{y}_T$  point from the
antiquark to the quark lines.}
\label{rbild}
\end{figure}
Inserting
(\ref{mchi2}) in (\ref{mcum}) we obtain
\begin{eqnarray}
  &&\Big\langle W_+^M(\frac{1}{2} \vec{b}_T,\vec{x}_T) \,
  W_-^M (-\frac{1}{2} \vec{b}_T,\vec{y}_T) \Big\rangle_G =
  \nonumber\\
  &&\; \; \; \frac{1}{9}\; {\rm{Tr}}_2 \; {\rm{exp}}
  [ (\frac{\lambda^a}{2} \otimes \frac{\lambda^a}{2}) \,
  (-i) \, \chi(\vec{x}_T,\vec{y}_T,\vec{b}_T)\;] .
  \label{mmatrix}
\end{eqnarray}
%
%
To evaluate the trace in (\ref{mmatrix})
we introduce two projectors $P_s$ and $P_a$
\begin{eqnarray}
  &&(P_s)_{(\alpha_1 \alpha_2)( \beta_1 \beta_2)} =
  \frac{1}{2}(\delta_{\alpha_1 \beta_1} \delta_{\alpha_2 \beta_2} +
  \delta_{\alpha_1 \beta_2} \delta_{\alpha_2 \beta_1}),
  \nonumber\\
  &&(P_a)_{(\alpha_1 \alpha_2)( \beta_1 \beta_2)} =
  \frac{1}{2}(\delta_{\alpha_1 \beta_1} \delta_{\alpha_2 \beta_2} -
  \delta_{\alpha_1 \beta_2} \delta_{\alpha_2 \beta_1})
  \label{projectors}
\end{eqnarray}
%
%
which act in the direct product space of two SU(3) quark
representations projecting onto the subspaces carrying
the irreducible representations. The decomposition is:
$3 \otimes 3 = 6 \oplus \bar{3}$. Using the identity:
\begin{eqnarray}
  &&\frac{\lambda^a}{2} \otimes \frac{\lambda^a}{2}=
  \frac{1}{3} P_s \;- \; \frac{2}{3} P_a ,
  \label{matrixcomp}
\end{eqnarray}
%
%
together with the projector
properties of $P_s$ and $P_a$ and  ${\rm{Tr}}_2 \, P_s=6$
and ${\rm{Tr}}_2 \, P_a=3$ we can immediately calculate the
trace in (\ref{mmatrix}) and so the colour dipole correlation
function:
\begin{eqnarray}
  &&{\rm Tr}_2 \exp \big[
  (\frac{\lambda^a}{2} \otimes \frac{\lambda^a}{2}) (-i\; \chi )
  \big]
  \nonumber\\
  &&={\rm Tr}_2 \exp \big[
  (\frac{1}{3} P_s \;- \; \frac{2}{3} P_a) (-i\; \chi )
  \big]
  \nonumber\\
  && ={\rm Tr}_2 \big[ P_s \, e^{-i{\T\, \frac{1}{3} \, \chi }}
  + P_a \, e^{i\, {\T\frac{2}{3} \, \chi}}   \big]
  \nonumber\\
  && =6\; e^{{-i \T\, \frac{1}{3} \, \chi }} +
  3\; e^{{i \T \, \frac{2}{3} \, \chi} } .
  \label{spurexp}
\end{eqnarray}
%
%

\subsection{The meson-meson amplitude}

Inserting (\ref{spurexp}) in (\ref{mamp})
and using  the Gaussian shaped
mesonic overlap functions from \cite{dfk}
our final result for the meson-meson
scattering amplitude  reads
\begin{eqnarray}
  &&T_{fi} = (2is) \, (2\, \pi)
  \int_0^{\infty} db \, b \, J_0(\sqrt{|t|}\, b)
  \hat{J}_{M,M}(b),
  \nonumber\\
  && \hat{J}_{M,M}(b)=
  -\int d^2x_T \,
    \frac{\T{ 1} }{ \T{ 2 \, \pi \, S_{H_1}^2} }
  \exp\left(-\frac{ \vec{x}_T^2}{2S^2_{H_1}}\right)   \,
        \int d^2y_T \,  {\T
    \frac{ \T{ 1 } }{ \T{ 2 \, \pi \, S_{H_2}^2 } } }
  \exp\left(-\frac{ \vec{y}_T^2 }{2S^2_{H_2}}\right)
  \nonumber\\[0.1cm]
  && \hphantom{\hat{J}_{M,M}(b)=} \Bigg\{ \;  \frac{2}{3}
  \cos (\frac{1}{3}
  \chi(\vec{x}_T,\vec{y}_T,\vec{b}_T)\,) +
  \frac{1}{3}
  \cos (\frac{2}{3}
  \chi(\vec{x}_T,\vec{y}_T,\vec{b}_T)\,) +
  \nonumber\\
  && \hphantom{\hat{J}_{M,M}(b)=} i \bigg[ -\; \frac{2}{3}
  \sin (\frac{1}{3}
  \chi(\vec{x}_T,\vec{y}_T,\vec{b}_T)\,)   \; +\frac{1}{3}
  \sin (\frac{2}{3}
  \chi(\vec{x}_T,\vec{y}_T,\vec{b}_T) \,)\; \bigg] - 1
  \,  \Bigg\} .
  \label{mresult}
\end{eqnarray}
%
%
Here $J_0$ is the zeroth-order Bessel function.
Due to (\ref{spurexp}) we get from the original
matrix valued exponential
(\ref{mcum}) a sum of two c-number valued
exponentials  which we have written
in terms of trigonometric functions.
Different hadrons are distinguished in
(\ref{mresult}) only through
their strong interaction
extension parameters $S_H$
which should be
of the order of the electromagnetic hadron radii.

Assuming $|\chi| \ll 1$ and
expanding $\hat{J}_{M,M}(b)$ in (\ref{mresult}) to the order
$O(\chi^2)$ gives the result of \cite{dfk}.
\begin{eqnarray}
  &&T_{fi} = (2is) \, (2\, \pi)
  \int_0^{\infty} db \, b \, J_0(\sqrt{|t|} \, b)
  \hat{J}_{M,M}^{(2)}(b),
  \nonumber\\
  && \hat{J}^{(2)}_{M,M}(b)=
  \int d^2x_T \,  {\T
    \frac{\T{ 1} }{ \T{ 2 \, \pi \, S_{H_1}^2} } }
  \exp\left(-\frac{ \vec{x}_T^2 }{2S^2_{H_1}}\right)
  \int d^2y_T \, {\T
    \frac{ \T{ 1} }{ \T{ 2 \, \pi \, S_{H_2}^2 } } }
  \exp\left(-\frac{ \vec{y}_T^2 }{2S^2_{H_1}}\right)  \nonumber\\
  && \hphantom{\hat{J}_{M,M}(b)=}
  \cdot \,\frac{1}{9} \; 
   (\chi(\vec x_T,\vec y_T,\vec b_T))^2, \qquad \qquad \qquad
   \qquad \qquad \qquad \; \,
  (|\chi| \ll 1 \,) .
  \label{mresultold}
\end{eqnarray}
%
%
But the integral $\hat{J}_{M,M}(b)$ in (\ref{mresultold})
is dominated by a region in $\vec x_T,\vec y_T$ where
$|\chi| \ll 1$ only for larger values of $b$,
say $b > 4 \,a$ (see section 4). So using
(\ref{mresultold}) $\hat{J}_{M,M}(b)$ can be calculated reliably
only for larger values of $b$.
Nevertheless with (\ref{mresultold}) the
total cross section and the slope parameter at $t=0$,
where one needs only the
first and third moments of
$\hat{J}_{M,M}(b)$,
were  calculated in \cite{dfk} in a satisfactory way.

Coming back to our expression (\ref{mresult}) for the elastic
scattering amplitude we show next that
this amplitude is purely imaginary.
Since $\chi$ is real, any real part of (\ref{mresult})
would have to come from the sine-terms. Now
the overlap functions in (\ref{mresult}) are invariant under
the replacements:
$\vec{x}_T \rightarrow -\vec{x}_T $ and $\vec y_T
\to -\vec y_T$, respectively, but
\begin{equation}
\chi (-\vec{x}_T, \vec{y}_T, \vec{b}_T)=
\chi (\vec{x}_T, -\vec{y}_T, \vec{b}_T)=
-\chi (\vec{x}_T, \vec{y}_T, \vec{b}_T) .
\label{chisym}
\end{equation}
%
%
This is easily seen from (\ref{mchi2}) since $\vec x_T\to -\vec x_T$
means the replacement $\vec r_{xq}\leftrightarrow r_{x\bar q}$
and $\vec y_T\to -\vec y_T$ the replacement $\vec r_{yq}
\leftrightarrow\vec r_{y\bar q}$.
Thus integrating over $\vec{x}_T$ and $ \vec{y}_T$ averages out the
real part in
(\ref{mresult}) and we get
\begin{eqnarray}
  &&T_{fi} = (2is) \, (2\, \pi)
  \int_0^{\infty} db \, b \, J_0(\sqrt{|t|}  \, b)
  \hat{J}_{M,M}(b),
  \nonumber\\
  && \hat{J}_{M,M}(b)=
  -\int d^2x_T \,  {\T
    \frac{\T{ 1} }{ \T{ 2 \, \pi \, S_{H_1}^2} } }
  \exp\left(-\frac{ \vec{x}_T^2 }{2S^2_{H_1}}\right)
  \int d^2y_T \,  {\T
    \frac{ \T{ 1} }{ \T{ 2 \, \pi \, S_{H_2}^2 } } }
  \exp\left(-\frac{ \vec{y}_T^2}{2S^2_{H_2}}\right)
  \nonumber\\[0.1cm]
  && \hphantom{\hat{J}_{M,M}(b)=} \Big[ \;  \frac{2}{3}
  \cos (\frac{1}{3}
  \chi(\vec{x}_T,\vec{y}_T,\vec{b}_T)\,) +
  \frac{1}{3}
  \cos (\frac{2}{3}
  \chi(\vec{x}_T,\vec{y}_T,\vec{b}_T)\,) -1  \Big].
  \label{mresult2}
\end{eqnarray}
%
%
As a consequence our meson-meson amplitude
is invariant
under the replacement of one hadron by its antihadron. The exchange of all
partons by their antipartons for a given parton configuration turns
around the loop direction. This results in a change of sign
of $\chi$ and so does not affect the amplitude (\ref{mresult2}).
In our approximations, we get only charge conjugation $C=+1$
(pomeron) exchange and no $C=-1$ (odderon) exchange contributions
to the amplitude.
A real part of the amplitude and $C=-1$ exchange contributions
could arise from higher cumulants in
(\ref{cumulantexpansion}).

We discuss now the constraints on the elastic amplitude implied by
the partial wave unitarity (see for instance \cite{31b}).
The partial wave expansion for the $T$ matrix element
for spin-zero mesons reads
\be
\label{29a}
T(s,t)=\frac{8\pi\sqrt s}{P_{cm}}\sum^\infty_{l=0}
(2l+1)P_l(\cos\vartheta)a_l(s),
\ee
%
%
\bear
\label{29b}
&&a_l(s)=\frac{1}{2i}(e^{2i\delta_l}\eta_l-1),\nonumber\\
&&0\leq\eta_l\leq1.
\ear
%
%
Here $P_l$ are the Legendre polynomials,
$P_{cm}$ is the cm momentum, $\vartheta$ the cm scattering
angle, and $\delta_l,\eta_l$ are the phase shifts and inelasticities,
respectively.

At high energies we get from (\ref{29a}) with $b=(2l+1)/\sqrt s$:
\be
\label{29c}
T(x,t)=8\pi s\int^\infty_0db bJ_0(b\sqrt{|t|})a_l(\sqrt s).
\ee
%
%
Comparison with (\ref{mresult2}) gives
\be
\label{29d}
\hat J_{MM}(b)=\left[-e^{2i\delta_l}\eta_l+1\right]
\Bigr|_{2l+1=b \, \sqrt{s} }.
\ee
%
%
Thus, the partial wave unitarity requires
\be
\label{29e}|\hat J_{MM}(b)-1|\leq1
\ee
%
%
and this is always satisfied for our amplitude (\ref{mresult2})
since $\chi$ is real and the profile functions are normalised
to one.

\section{Proton-proton and proton-antiproton scattering in the
            quark-diquark picture}

A lot of experimental data is available for elastic differential cross
sections of proton-proton ($pp$) scattering
up to $\sqrt{s}=63$ GeV and proton-antiproton
($p\bar{p}$) scattering up to Tevatron  energies of $\sqrt{s}=1800$ GeV
\cite{ppdat1,ppdat2,ppdat3}.
In this section we will compare the high energy data in
the range $\sqrt{s} \geq 23\ GeV$ to the results of our calculations
making the assumption that the proton has a quark-diquark
structure. Thus we can use the formulae for the
meson-meson amplitude as presented in Sect. 3.

Our starting point is (\ref{mresult2}) depending on 4 parameters:
the QCD vacuum parameters $G_2, \ \kappa$ and $a$ and the proton
extension parameter $S_{H_1}=S_{H_2}=S_p$. The vacuum
parameters are surely energy and process independent,
the extension parameter $S_p$ will be allowed
to vary with energy. The numerical calculations
using (\ref{mresult2}) are too lengthy to attempt a ``best fit'' of
these parameters from the data. We adopted the following procedure
instead. In the SVM with the ansatz 
(\ref{corfktmom1},\ref{corfktmom}) for the functions
$D,D_1$ the string tension $\rho$ is given by
\bear
\label{29}
\rho&=&\frac{\pi^3\kappa G_2}{36}\int^\infty_0 dZ^2D(-Z^2)\nonumber\\
&=&\frac{32\pi\kappa G_2a^2}{81}.
\ear
%
%
Typical values for $\rho$ extracted from phenomenology
(cf. e.g. \cite{26a}) are
$ \rho=(420\pm 20\ MeV)^2 $. In the following we will
express $G_2$ through $\rho$ and $a$ using (\ref{29}).
From previous work \cite{dfk,mdoc} and the lattice measurements
discussed in Sect. 3 we expect for the correlation length
0.30\ fm $\lsim\ a\ \lsim\ 0.37$\ fm, for $\kappa\approx 0.75$ and
for the proton extension parameter at $\sqrt s=23$ GeV 
$S_p\approx 0.86$
corresponding to the electromagnetic proton radius. Now we
considered again $\sqrt s=23$ GeV and started our numerical
investigations using (\ref{mresult2}), 
varying the parameters
$\rho, a, \kappa, S_p$ aroung the values indicated above. We
calculated
\be
\label{37x}
\frac{d\sigma}{dt}=\frac{1}{16\pi}
\frac{1}{s^2}|T_{fi}|^2
\ee
%
%
and the total cross section, using
the optical theorem which reads for $s\gg m^2_p$:
\be
\label{33}
\sigma_T(pp)=\frac{1}{s} {\rm Im} (T_{fi})
\ee
%
%
and compared to experiment.
The experimental data on the total $pp$
and $p\bar p$ cross sections is very well described by the DL fit
\cite{dola1}, \cite{particle}. We are only interested in the pomeron
exchange part here. Therefore we take as ``experimental''
input the pomeron part of $\sigma_T(pp)$ in the DL
parametrization \cite{dola1}:
\be
\label{pompart}
\sigma_T(pp)\Big|_{\exp}=21.7\left(\frac{s}
{{\rm GeV}^2}\right)^{0.0808}\ {\rm mb}.
\ee
%
%
We imposed as constraint that our amplitude reproduced (\ref{pompart})
exactly at $\sqrt s=23$ GeV.

With this procedure we found quite a satisfactory description
of the data for $d\sigma/dt$ as shown in Fig. 5 for the following
values of our parameters:
\be
\label{31}
\rho=(435\ {\rm MeV})^2,
\ee
%
%
%
%
\be
\label{32}
a=0.32\ {\rm fm},
\ee
%
%
%
\be
\label{32a}
\kappa=0.74,
\ee
%
%
%
\be
\label{32b}
S_p(s=(23\ {\rm GeV})^2)=0.87\ {\rm fm}.
\ee
%
%
%
%
From (\ref{29}), (\ref{31}), (\ref{32}) we get
\bear
\label{34a}
G_2&=&(529\ {\rm MeV})^4,
\nonumber\\
\kappa G_2 a^4&=&0.40.
\ear
%
%
%
Varying the parameters away from the values
(\ref{31}-\ref{32b}) did not lead to improvements. Also, the values  
(\ref{31}-\ref{32b}) are well within
the range obtained in \cite{dfk,mdoc} and quite compatible with the
lattice results (\ref{10a}). In the following we will thus
fix the vacuum parameters $\rho, a,\kappa$
to their values (\ref{31}-\ref{32a}).

Now we consider $\sigma_T$ and $d\sigma/dt$ at
higher energies $\sqrt s$, where we have then only one free
parameter $S_p(s)$ left. We fix $S_p(s)$ again by
requiring that our model reproduces the experimental
value for the pomeron part of $\sigma_T(pp)$ according
to (\ref{pompart}).

On the other hand we follow \cite{dfk}, \cite{15b} and
fit our calculated values of $\sigma_T(pp)$ from
(\ref{mresult2}), (\ref{33}) as shown in
Fig. \ref{sigplot} 
%
%
%
\begin{figure}[htb]
 
\vspace*{-1.0cm}
 
\hspace{1cm}
\epsfysize=8.5cm
\centerline{\epsffile{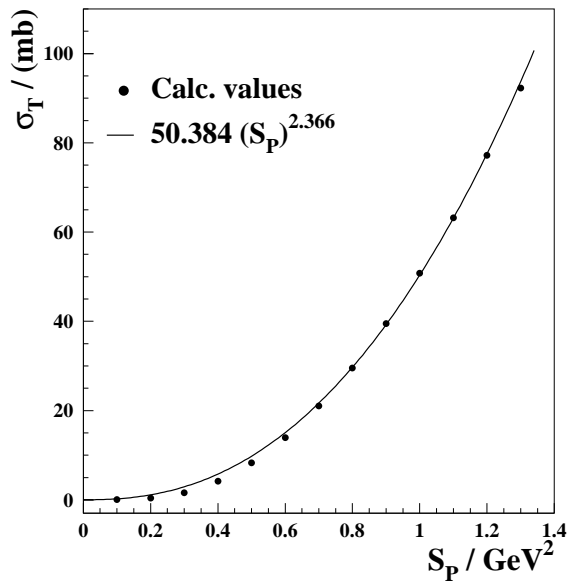}}
 
\vspace*{0cm}
 
\caption[a]{ \em The total cross section 
$\sigma_T$ 
versus $S_P$. The solid points are the calculated
values,
the solid curve corresponds to the
power fit.  }
\label{sigplot}
\end{figure}
%
%
%
to a power of $S_p$. We get
a good description in the range $2.5 \, a \leq S_p \leq 4.0 \, a$ 
with
\be
  \label{34}
  \sigma_T(S_p)\Big|_{cal}=50.384 
  \left (S_p \right)^{2.366}
  \, {\rm mb}.
\ee
%
%
%
Equating (\ref{33}) to (\ref{34}) we get $S_p$ as function of
$s$:
\bear
  \label{expar}
  S_p(s) = 0.700 \, 
  \left(\frac{s}{{\rm GeV}^2}\right)^{0.034}
  \, {\rm fm}.   
\ear
%
%
This leads to $S_p=0.86,\ 0.87,\ 0.93,\ 1.07$
and $1.17\ {\rm fm}$ for $\sqrt s=20,\ 23,\ 63,\ 546$ and 1800 GeV,
respectively.
%
%
%
\begin{figure}[htb]
 
\vspace*{-1.0cm}
 
\hspace{1cm}
\epsfysize=14cm
\centerline{\epsffile{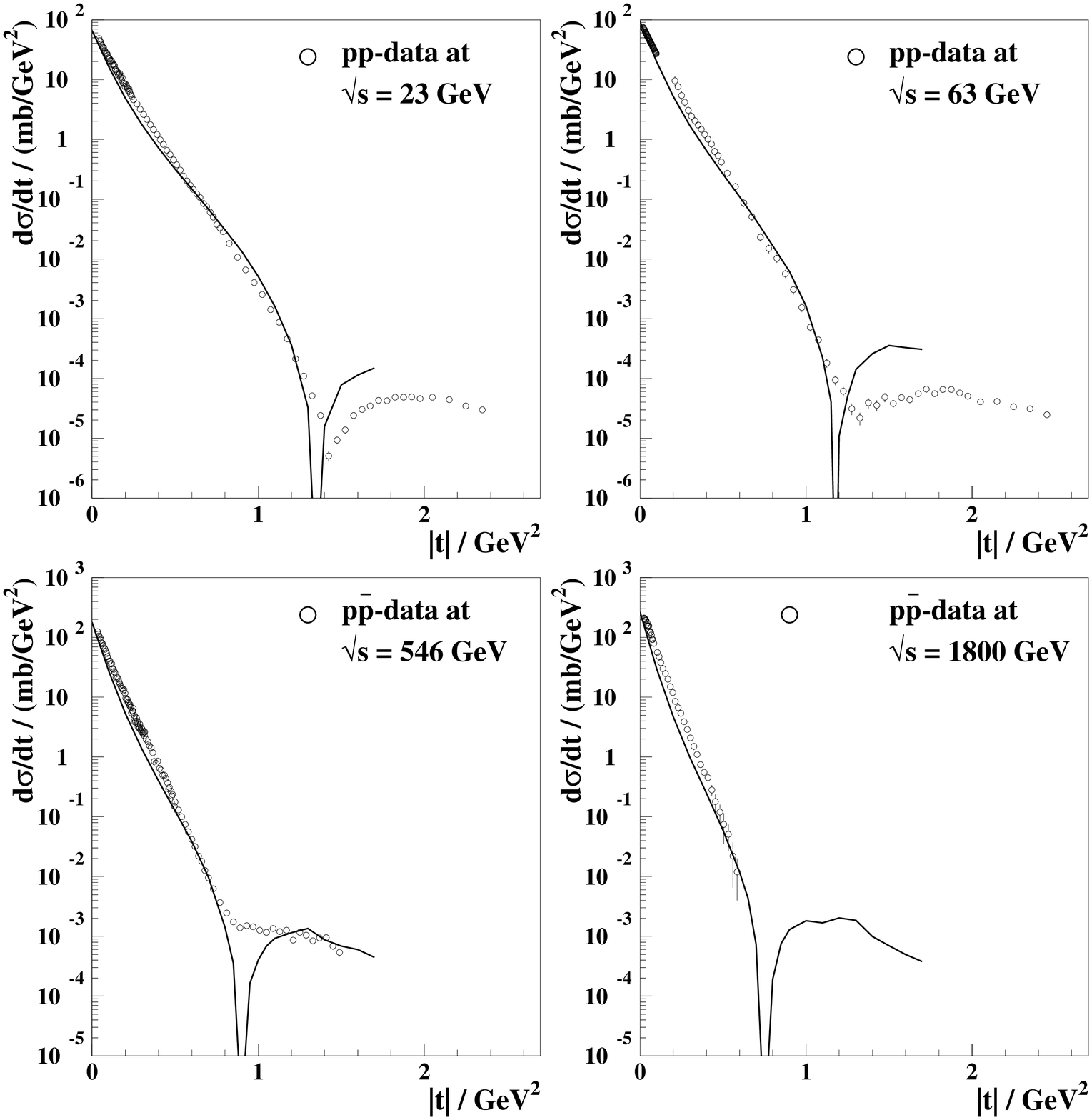}}
 
\vspace*{0cm}
 
\caption[a]{ \em Differential elastic cross sections for
c.m. energies $\sqrt{s}=23,63,546$ and $1800${\rm{ GeV}}.
This corresponds to proton extension parameters
$S_P=0.87, \, 0.93, \, 1.07$ and $1.17$ {\rm{fm}}.
The data
at $\sqrt{s}=23$ and $63$ {\rm{GeV}} are from the ISR
experiment {\rm{\cite{ppdat1}}}. The $p\bar{p}$ scattering data
at $\sqrt{s}=546$ GeV are from the  
{\rm{\cite{ppdat2}}} and the data at $\sqrt{s}=1800$ GeV
from {\rm{\cite{ppdat3}}}.}

\label{probild}
\end{figure}
%
%
%
%
For comparison, the mean squared charge radius of 
the proton as determined from  Lamb-shift
measurements \cite{radii1} is $r_P=0.89 \pm 0.014 {\rm fm}$.
Now everything is fixed and we can compute the elastic
scattering amplitude and $\sigma_T$
from (\ref{mresult2}). In Fig. \ref{probild} we present our results.
A first observation is that for all energies
the calculated differential
distributions follow the experimental data quite well over
many orders of magnitude. The fact that this is true  up
to  $\sqrt{s} = 1800\, GeV$ supports
the description of the s-dependence by a
$s$-dependent extension parameter $S_p(s)$.

For small $|t|\simeq 0.25$ GeV$^2$
we get a change of slope
(Fig. \ref{smallt}).
%
%
%
\begin{figure}[htb]
 
\vspace*{-1.0cm}
 
\hspace{1cm}
\epsfysize=9cm
\centerline{\epsffile{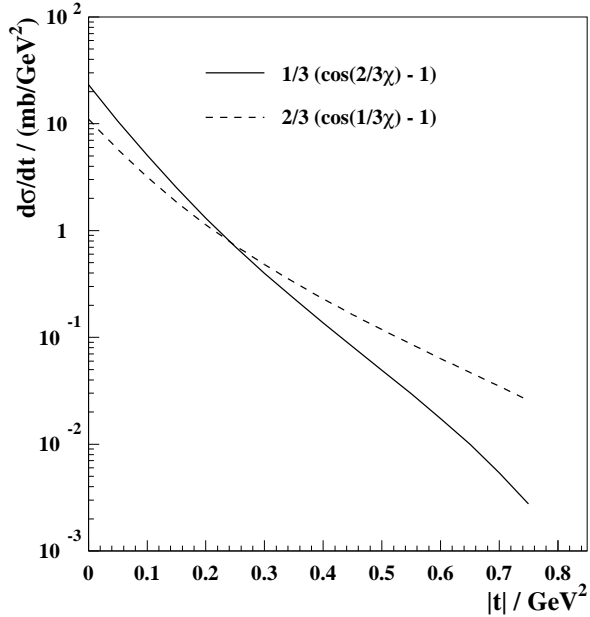}}
 
\vspace*{0cm}
 
\caption[a]{ \em
Differential elastic cross sections
calculated for the
values of $G_2= (529 {\rm MeV})^4$,
$a=0.32 \ {\rm fm} $ and $ S_p=0.87 \ {\rm fm} $
using as integrand of the amplitude {\rm{ (\ref{mresult2})}}
$\frac{1}{3} [\cos(\,\frac{2}{3} \, \chi)-1]$
(solid line) and
$\frac{2}{3}[\cos(\,\frac{1}{3} \, \chi)-1]$
(dashed line) respectively.} 

\label{smallt}
\end{figure}
%
%
Splitting the integrand of
(\ref{mresult2}) into two contributions
\be
\label{37}
\frac{2}{3}\cos(\frac{1}{3}\chi)+\frac{1}{3}\cos(\frac{2}{3}
\chi)-1=
\frac{1}{3}\left[\cos(\,\frac{2}{3} \, \chi)  
-1\right]+\frac{2}{3}\left[\cos(\,\frac{1}{3} \,
\chi )-1\right]
\ee
%
%
we find that for $|t| \,
{\mathrel{\vcenter
    {\hbox{$<$}\nointerlineskip\hbox{$\sim$}}}}
\, 0.25 $ GeV the
first term dominates and for $|t|\gsim 0.25$ GeV the second one dominates.
Such a change of slope is indeed reported by experiments \cite{31a}.

For all energies the imaginary part of our amplitude changes sign
at some $t<0$. Due to the absence of a real part in
(\ref{mresult2}) the calculated
differential cross sections have a zero there.
This causes an infinitely deep dip in our $t$-distributions.
We expect this dip to be at least partly filled up
once we change to more general quark
configurations and include higher cumulant terms.
The point at which the zero occurs in our calculation
moves to smaller values of
$|t|$ with increasing energy and is always in the region where
experiments see a marked structure: At lower energies there is
a dip in $pp$, and a shoulder in $p\bar p$ scattering,
respectively. At the highest energies only $p\bar p$ data is
available and one finds a shoulder. Thus our model
produces structure in the $t$-distributions at the right place.
But we should insert the warning that
these dips occur at $|t| \simeq 1$ GeV$^2$
where one would expect also
perturbative effects to play a significant role.
(cf. \cite{laper},\cite{dolaper}).

Of course, our model does not give a perfect fit to the data.
At all energies our calculated curves are somewhat too steep
at very small $|t|$. Our amplitude is purely imaginary and
thus does not satisfy the relation between the phase and
the $s$-dependence required by analyticity
and Regge theory \cite{regge}. Also our
$d\sigma/dt$ is the same for $pp$ and $p\bar p$ scattering, whereas
experimentally these differ markedly in the dip region.
This was nicely explained theoretically in \cite{dolaper}
as an interference of single and double pomeron
and three-gluon exchange. We will
have to see if higher cumulant terms and/or a departure from
the strict quark-diquark picture of the proton
will lead us to an improvement on these points in our model.

In Fig. \ref{lhc} we show our prediction for $d\sigma/dt$ in $pp$ scattering
at $\sqrt s=14$ TeV corresponding to the LHC energy. 
%
%
%
\begin{figure}[htb]
 
\vspace*{-1.0cm}
 
\hspace{1cm}
\epsfysize=10cm
\centerline{\epsffile{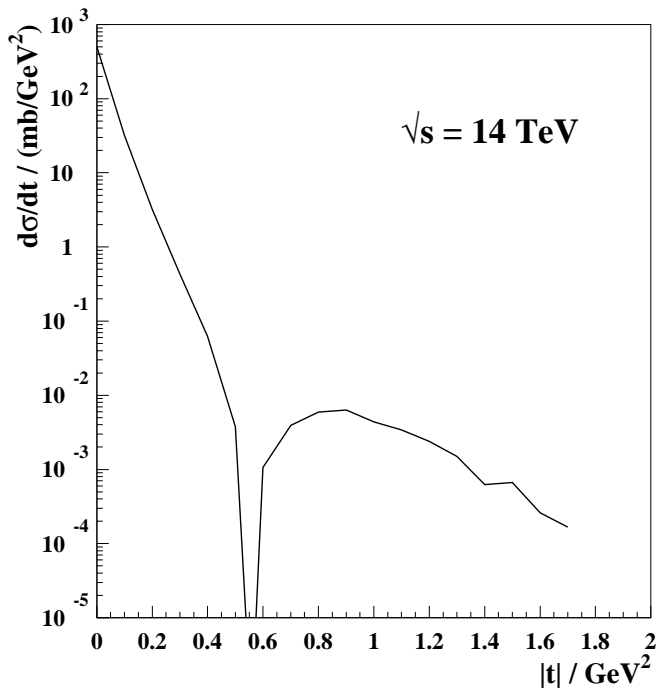}}
 
\vspace*{0cm}
 
\caption[a]{\em Prediction of the differential
elastic cross section for the
LHC energie  $\sqrt{s}=14 \, {\rm{TeV}}.$

 }
\label{lhc}
\end{figure}
%
%
%
%
The total
cross section is again assumed to be given by the DL
parameterisation (\ref{pompart}). This was used as constraint
to fix the extension parameter $S_p$ according to Fig. 4
with the result
\be
\label{38a}
S_p(s=(14\ {\rm TeV})^2)=1.34 \, {\rm fm}.
\ee
%
%
But the shape of $d\sigma/dt$ is a prediction of our model
and we are looking forward to the corresponding
experimental data.

We will show next that our results for $d\sigma/dt$ depend
crucially on the string tension $\rho\not=0$, i.e. on the
confinement features of QCD. For this we plot in
Fig. \ref{kapdsig} 
$d\sigma/dt$ for $\sqrt s = 23$ GeV calculated for the
same values of $G_2= (529 {\rm MeV})^4$,
$a=0.32 \ {\rm fm} $ and $ S_p=0.87 \ {\rm fm} $, but
for different values of $\kappa$.
%
%
%
\begin{figure}[htb]
 
\vspace*{-1.0cm}
 
\hspace{1cm}
\epsfysize=10cm
\centerline{\epsffile{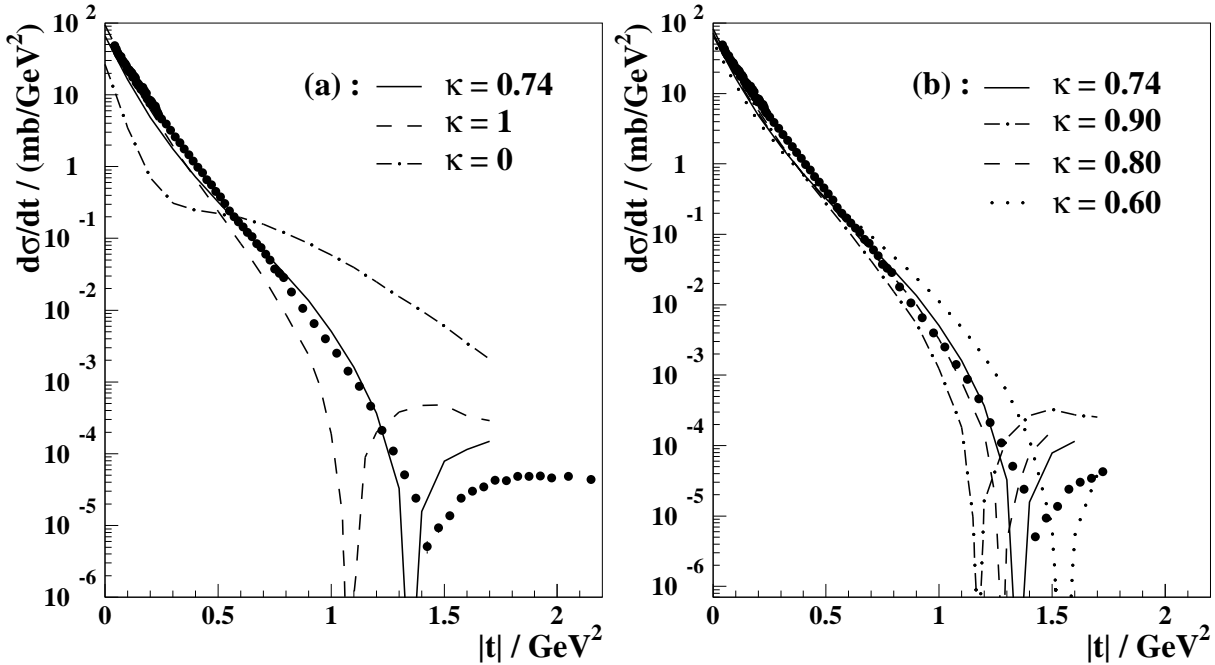}}
 
\vspace*{0cm}
 
\caption[a]{\em
The variation of the differential elastic
cross section 
calculated for the
same values of $G_2= (529 {\rm MeV})^4$,
$a=0.32 \ {\rm fm} $ and $ S_p=0.87 \ {\rm fm} $;
{\rm (a)} for $\kappa=0.74$ (solid line), 
$\kappa = 1$ (dashed line) and 
$\kappa=0$ (dash-dotted line);
{\rm (b)} for 
$\kappa=0.74$ (solid line),
$\kappa=0.90$ (dash-dotted line),
$\kappa=0.80$ (dashed line) and
$\kappa=0.60$ (dotted line) 
together with the experimental data
(solid points)
at $\sqrt{s}=23$ {\rm GeV}. From {\rm (b)} one can
see, that a value of  $\kappa$
around $0.74$ is favoured by the data.}
\label{kapdsig}
\end{figure}
%
%
%
%
For the purely non-abelian
case $\kappa=1$ the fall of $d\sigma/dt$ is too steep, for the
purely abelian, the non-confining case $\kappa=0$, the
fall is much
too slow with increasing $|t|$. The correct fall
and dip position is obtained
for $\kappa = 0.74$, not far from the value determined from the lattice
calculations (\ref{10a}).

To explain this we plot the profile function
$\hat{J}_{M,M}(b)$
multiplied with $b$ for $\kappa=0$ and
$\kappa = 1 $ versus $b$ in Fig. \ref{jkappa}.
%
%
%
%
\begin{figure}[htb]
 
\vspace*{-1.0cm}
 
\hspace{1cm}
\epsfysize=9cm
\centerline{\epsffile{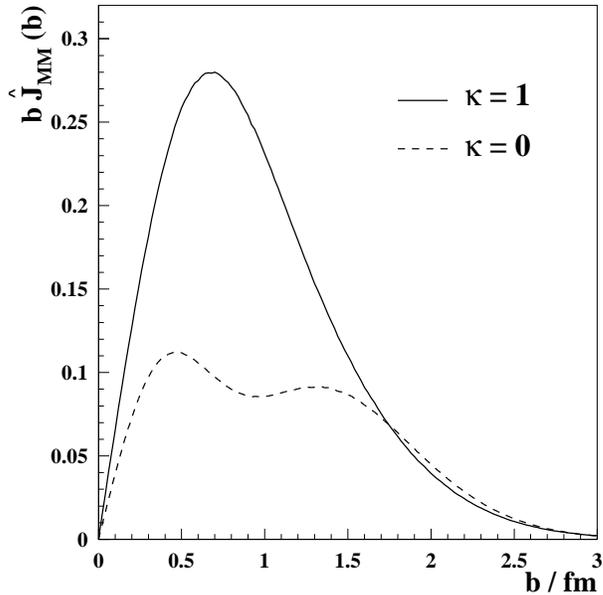}}
 
\vspace*{0cm}
 
\caption[a]{ \em
The dimensionless profile function 
$\hat{J}_{MM}(b)$ times $b$ versus $b$ 
for $G_2= (529 {\rm MeV})^4$,
$a=0.32 \ {\rm fm} $ and $ S_p=0.87 \ {\rm fm} $
in the cases $\kappa=1$ (solid line) and
$\kappa=0$ (dashed line).}
\label{jkappa}
\end{figure}
%
%
%
%
As we can see,  the two curves differ noticeably.
For $\kappa = 1 $ we get a single maxima
whereas for $\kappa = 0 $ we get two maxima.
This can be understood
by looking at the dependence of $\chi$ on $b$ for fixed
transverse vectors $\vec{x}_T$ and $\vec{y}_T$ in the two cases.
As we see from (\ref{mchi2}) $\chi$ is a sum of four terms (replace
here $\bar q$ by the diquark $qq$) corresponding to
$q-q,\ q-qq$ etc. interactions. For $\kappa=0$
the function $I(\vec r_x,\vec r_y)$ in (\ref{mchi}) depends
only on the difference $\vec r_y-\vec r_x$ of the parton positions,
but for $\kappa=1$ $I(\vec r_x,\vec r_y)$
gets contributions from all ``strings'' spanned
between $o$ and the $q$'s and $qq$'s positions ( Fig. \ref{chiplot}).
%
%
%
%
\begin{figure}[htb]
  \unitlength1.0cm
  \begin{center}
    \begin{picture}(15.,8.8)

      \put(-0.3,7.2){({\large a}):}      

      \put(12.0,7.2){({\large b}):}

      \put(-0.4,3.0){
        \epsfysize=5.5cm

        \epsffile{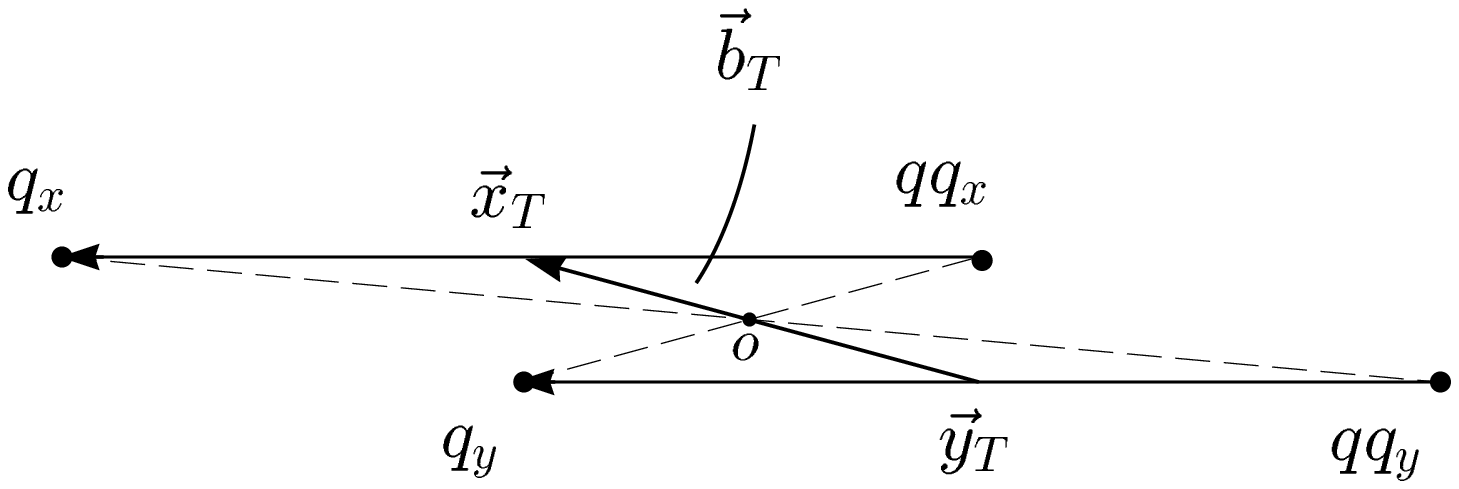}}
      \put(8.3,0.0){
        \epsfysize=9.4cm

        \epsffile{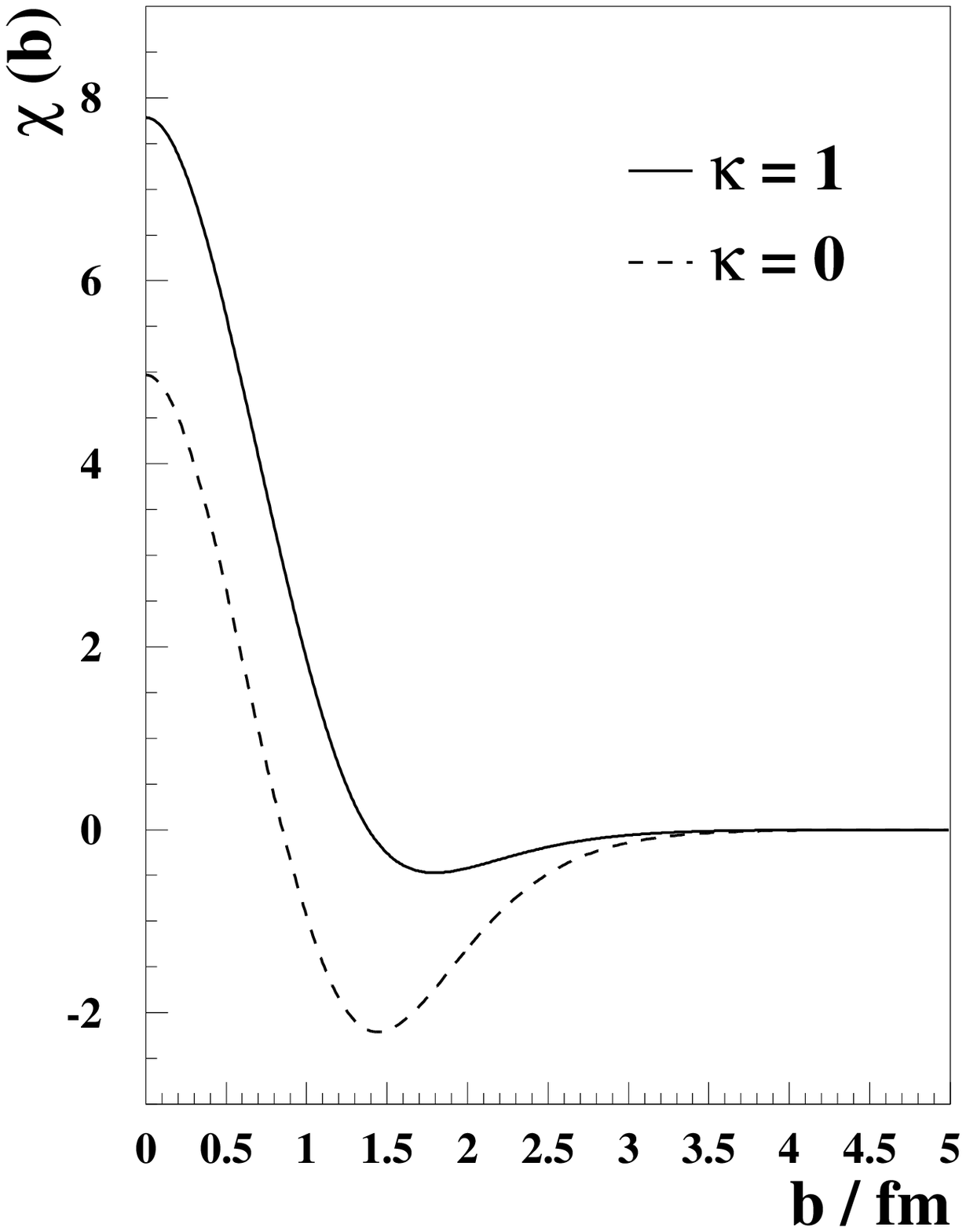}}

    \end{picture}
  \end{center}
\vspace*{-0.5cm}
\caption{\em {\rm (a)} The transverse 
  positions of the quarks $q_x, \, q_y$
  and diquarks $qq_x, \, qq_y$ for parallel 
  loops. The dashed lines represent the 
  ``strings'' spanned between o and the $q$'s 
  and $qq$'s positions.
  {\rm (b)} The correlation function $\chi(b)$
  versus b for
  $|\vec{x}_T|=|\vec{y}_T|=4\, a$ and
  $\vec{x}_T ,\  \vec{y}_T,\
  \vec{b}_T$  parallel to each other
  in the cases $\kappa=1$ (solid line) and 
  $\kappa=0$ (dashed line). The curves 
  correspond to  loop configurations
  sketched
  in {\rm (a)}, namely to
  parallel loops lying on each 
  other for $b=0$ and moving 
  for  increasing  $b$  in $+\vec{x}_T$
  direction
  and $-\vec{y}_T$ direction 
  respectively. }
  \label{chiplot}
\end{figure}
%
%
%
%
We get large
contributions to $I$ if the arguments of the Bessel functions
$K_{2,3}$ in (\ref{mchi}) are small. For $\kappa=0$ this means
that the partons $(q,qq)$ have to be close in transverse space,
but for $\kappa=1$ it is only required that one parton be close
to the string of the other parton. To see the
consequences of this,
we take as an example
$|\vec{x}_T|=|\vec{y}_T|=4$ and
$\vec{x}_T ,\  \vec{y}_T,\
 \vec{b}_T$ nearly parallel to each other
(Fig. \ref{chiplot}). For $\kappa=1$
we have the following situation.
For small and
medium values of $b$ the terms $I$
corresponding to the $q_x-q_y$ and $qq_x-qq_y$ 
interaction dominate in the
sum for $\chi$ in (\ref{mchi2}). 
These functions $I$
decrease with increasing $b $
nearly monotonously corresponding to $q_x$ moving away from the
string $o-q_y$ and $qq_y$ away from the string of $o-qq_x$.
But the interaction of $q_y$  with the string $o-q_x$
and  $qq_x$ with the string
$o-qq_y$ remain dominant. 
Compared to them the terms $I$ corresponding to 
the $qq_x-q_y$ which enters with negative sign in 
(\ref{mchi2}) stays smaller and also the function
$\chi$ decreases nearly monotonously with 
increasing $b$. In summary: 
for $\kappa=1$ the function $\chi$ 
is sizable and practically  always of the same sign
as long as partons are
close to strings of other partons, i.e. as long as projection
of the two loops overlap.
This picture,
smeared out due to the integrations over the transverse
vectors, is reflected in the shape of
$(b \, \hat{J}_{M,M}(b))$ for $\kappa=1$.

For $\kappa=0$ the function $\chi$ has a maximum at $b=0$,
where the quarks and the
diquarks of the two protons are closest to
each other in transverse space. Then $\chi$ decreases rapidly for
increasing $b $  on a scale given by the correlation length
$a$, passes zero at some value $b_0$
and approaches a minimum. The latter corresponds to the situation
where the  quark
of one proton is very close to the diquark of the other proton.
Since there are no strings in this case all interactions
except the one between $qq_x$ and $q_y$ are
then negligeable and the latter enters with opposite sign
to the $q_x,q_y$ and $qq_x,qq_y$ interaction in (\ref{mchi2}).
For $b$ increasing further
the transverse distance between all partons of the dipoles increase
and $\chi$ goes to zero. At $b_0$ the expression in square
brackets in (\ref{mresult2}) vanishes. Thus we expect  --
after smearing out through the integrations -- to see a minimum
in $b\hat J_{MM}(b)$ around $b_0$ and this is indeed seen
in Fig. \ref{jkappa}. Thus, through this sign change of $\chi$
we understand the two maxima in the
shape of $b\, \hat{J}_{M,M}(b)$ for $\kappa = 0$,
and also that for $\kappa=0$  the first maximum
in $b\hat J_{MM}(b)$ occurs  for a smaller value
of $b$ than for $\kappa=1$. After the Fourier-Bessel transformation
in (\ref{mresult2}) this translates immediately in the slower
decrease of $d\sigma/dt$ for $\kappa=0$ compared to $\kappa=1$.
The dip in $d\sigma/dt$ is generated by a cancellation of
positive and negative contributions from $b\hat J_{MM}(b)J_0(ba
\sqrt{|t|})$. For $\kappa=1$ and $\kappa=0.74$ the function
$b\hat J_{MM}(b)$ is ``smooth'' and the oscillating Bessel
function brings the integral (\ref{mresult2}) ``easily''
to zero for some $|t|$. For $\kappa=0$ the ``oscillation''
of $b\hat J_{MM}(b)$ together  with the
oscillation of $J_0(ba\sqrt{|t|})$ can produce a dip  only
at much higher $|t|$. From this point of view
the structure of the $pp$ scattering amplitude gives us
direct information that a string formation between
quarks in QCD is essential also here.

We can now also understand how the
shrinkage of our  calculated differential cross sections (see
Fig. \ref{probild}) for increasing c.m. energies is caused:
We found above for $\kappa=0.74$ that the shape of
$d \sigma / d t$ is controlled by string scattering.
We assumed $S_p$ to increase with $s$ which means increasing
lengths of the string sizes and thus still
overlap for
larger values of the impact parameter $b$. This in turn
translates into steeper forward peaks
of the scattering amplitude and so into steeper elastic differential
cross sections at higher energies.
%
%
%
\begin{figure}[htb]
 
\vspace*{-1.0cm}
 
\hspace{1cm}
\epsfysize=9cm
\centerline{\epsffile{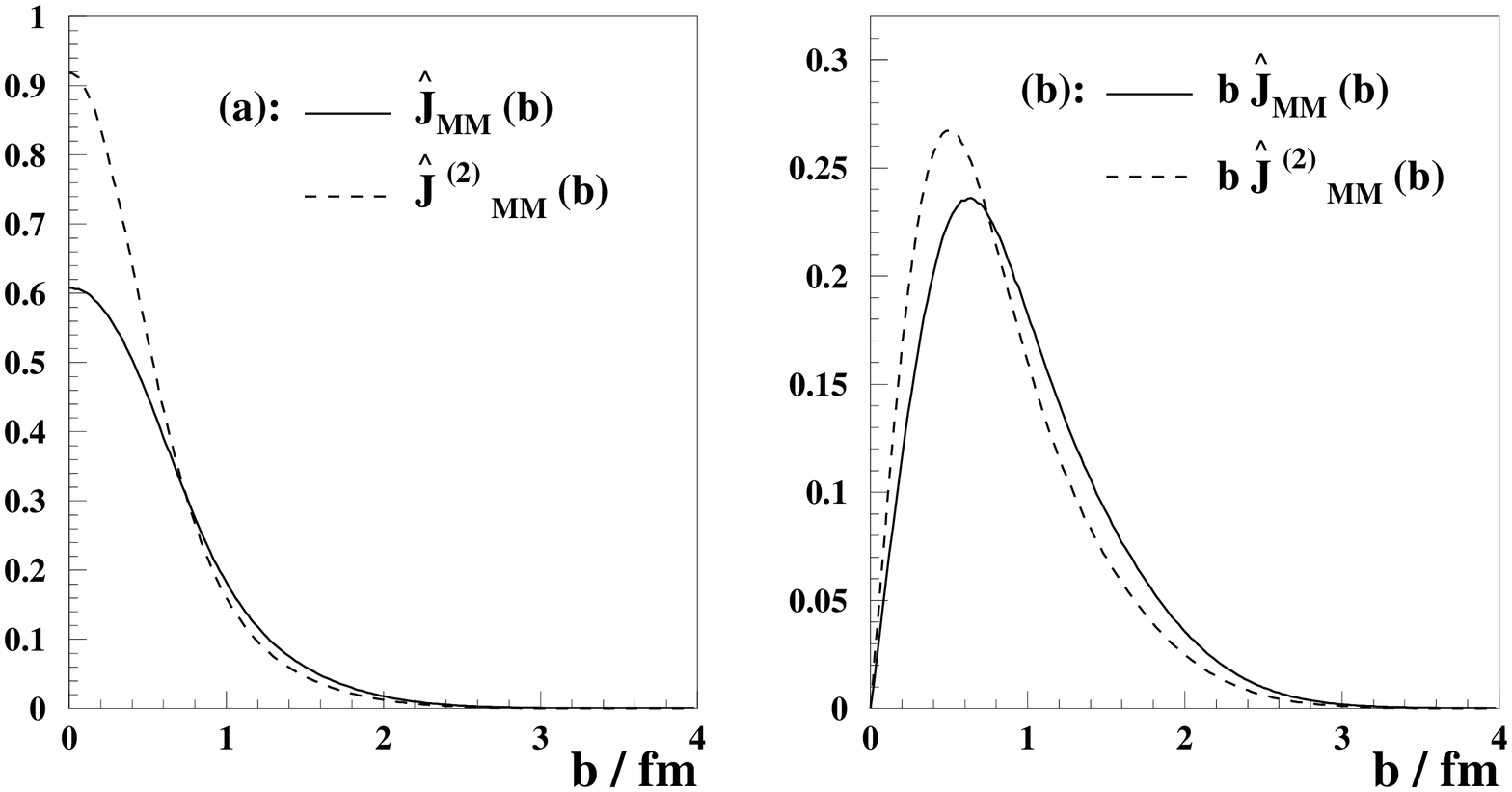}}
 
\vspace*{0cm}
 
\caption[a]{\em

  For $\sqrt{s}=23$ {\rm GeV}
  {\rm (a)} the profile functions $\hat{J}_{MM}(b)$ (solid line)
  and $\hat{J}_{MM}^{(2)}(b)$ (dashed line) as a function
  of  $b$.
  {\rm (b)} The profile functions $\hat{J}_{MM}(b)$ (solid line)
  and $\hat{J}_{MM}^{(2)}(b)$ (dashed line) 
  multiplied with $b$  versus $b$. }
\label{uni}
\end{figure}
%
%

In Fig. \ref{uni} we compare
our profile function $\hat J_{MM}(b)$
from (\ref{mresult2}) with the corresponding one, $J_{MM}^{(2)}(b)$
of (\ref{mresultold}),
where only the term of order $\chi^2$ is kept. We see
from Fig. \ref{chiplot} that this can only be a good approximation for larger
values of $b$ and indeed, the two functions $\hat J_{MM}$ and
$\hat J^{(2)}_{MM}$ are quite similar for $b\gsim 2a$, but are
rather different for $b\lsim 2a$. But for the total cross section
and the slope parameter at $t=0$ which are obtained from integrals
over $b \hat J_{MM}(b)$  and $b^2\hat J_{MM}(b)$ this does
not make much difference. On the other hand at larger and larger
$|t|$ the integrals with the Bessel function in (\ref{mresult2}),
(\ref{mresultold})
probes smaller and smaller vales of $b$ in $\hat J_{MM}(b)$ and then
the differences between $\hat J_{MM}$ and $\hat J_{MM}^{(2)}$
show up clearly.
%
%
\begin{figure}[htb]
 
\vspace*{-0.2cm}
 
\hspace{1cm}
\epsfysize=6cm
\centerline{\epsffile{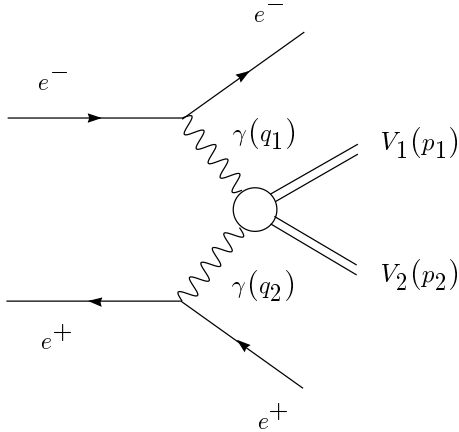}}
 
\vspace*{0cm}
 
\caption[a]{ \em Vector meson production 
in two photon processes at $e^+e^-$ colliders.  }
\label{feyn}
\end{figure}
%
%
%

Finally we come back to the partial wave expansion
for the T matrix element
at high energies (\ref{29c}) and 
discuss the case of total absorption which means
$\eta_l=0$ for the inelasticities $\eta_l$ in (\ref{29d}).
Thus total absorption requires 
$|\hat{J}_{MM}(b)|\le 1$ which is again satisfied
for our amplitude (\ref{mresult2}). At 
$\sqrt{s}=23$ GeV this can be read off from
Fig. \ref{uni}a.

\section{Meson-meson and meson-baryon scattering}
\subsection{Meson-meson scattering}

Meson-meson scattering is the reaction best suited to apply our
formula (\ref{mresult2}).
Unfortunately  there is no data available for small $t$ elastic
meson-meson-scattering at sufficiently large c.m. energies. But the
reaction
\begin{equation}
\gamma(q_1) + \gamma(q_2) \rightarrow V_1(p_3) + V_2(p_4) , \;
 \; \; \;  V_{1,2}=\rho, \omega, \phi
\label{reaction}
\end{equation}
%
%
is nearly as good to extract mesonic differential cross
sections as a purely mesonic one and it can be
studied for instance
at LEP \cite{gamma} in photon-photon processes (Fig. \ref{feyn}).
Certainly at least part of the reaction (\ref{reaction})
is due to the vector dominance model
\cite{vmd} mechanism:
The photons fluctuate before the collision into vector mesons
which then  interact.
The hadronic interaction can be calculated in the
way described in Sect. 3. Thus in essence we can regard the
reaction (\ref{reaction}) as elastic scattering of
vector mesons.
This approximation should be best for quasi real photons
in (\ref{reaction}), i.e. for very small virtualities $|q_1^2|$ and
$|q^2_2|$.

In this spirit we discuss now $\rho\rho$ elastic scattering as an
example. In our amplitude (\ref{mresult2}) we have already fixed
the vacuum parameters $\rho,\kappa$ and $a$ in Sect. 4, but we
still have to choose the $\rho$ extension parameter $S_\rho$
and we assume it to be equal to the $\pi$ extension 
parameter as determined in the next section

Indeed, data on $\rho$-$N$ total cross sections from
photoproduction on nuclei at low energies \cite{400} show
$\sigma_T(\rho N) \approx \sigma_T(\pi N)$.
Thus we assume:
\be
\label{38}
S_\rho=0.60\ {\rm fm}\quad {\rm for}\quad \sqrt s=20\ {\rm GeV}.
\ee
%
%
Now everything is fixed and we can calculate $d\sigma/dt$ for
$\rho$-$\rho$ scattering.

In Fig. \ref{rhobild} we plot  $d \sigma/d t$ normalised
to 1 at $t=0$.
%
%
%
\begin{figure}[htb]
 
\vspace*{-1.0cm}
 
\hspace{1cm}
\epsfysize=9.5cm
\centerline{\epsffile{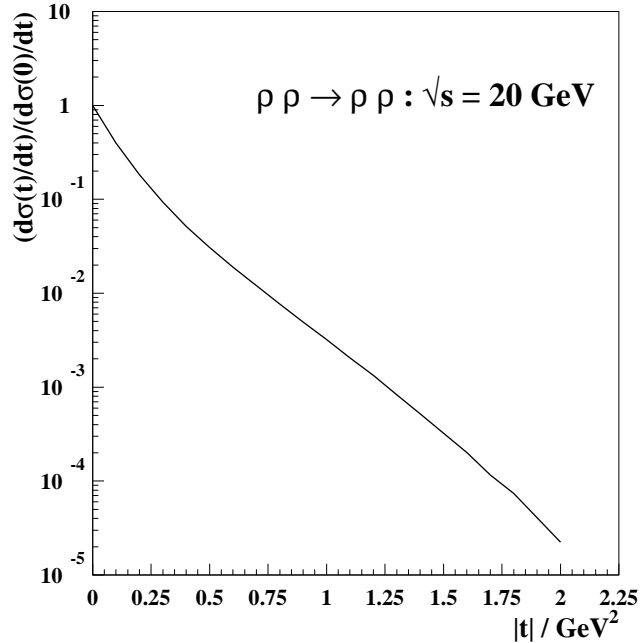}}
 
\vspace*{0cm}
 
\caption[a]{\em
Prediction for the differential
elastic cross section of $\rho \rho$-scattering
at   $\sqrt{s}=20$ {\rm{GeV}}
using $S_{\rho}=0.60$ {\rm{fm}}. }
\label{rhobild}
\end{figure}
%
%
%
%
Assuming the vector dominance model and
neglecting contributions from non-diagonal scattering
$\omega\omega\to\rho\rho$ etc., Fig. \ref{rhobild} also  gives the shape
of $d\sigma /d t$ for
$\gamma \gamma \rightarrow \rho \rho$-scattering.
It would be very interesting to have data for instance
from LEP2 to compare with our $t$-distribution.

\subsection{Differential cross sections for meson-proton scattering}

Here we calculate elastic differential cross sections
for $\pi^{\pm} p$ and  $K^{\pm} p$ scattering at $\sqrt{s}=19.5 \, GeV$.
This is the largest energy
for which data for these reactions exist.

In our scattering amplitude (\ref{mresult2}) we have already fixed
vacuum parameters $\rho,\kappa, a$ (\ref{31}-\ref{32a}). We
consider the proton as a quark-diquark system with
extension parameter $S_p$ and
use $S_p=0.86$ fm from (\ref{expar}). To fix the meson extension
parameters we again normalise
our total cross sections to the pomeron parts in the
DL parametrisations of the  $\pi^{\pm} p$ and $K^{\pm} p$
total cross sections \cite{dola1}.
\bear
\label{40}
&&\sigma(\pi^\pm p)|_{Pom.}= 22.0 \, {\rm{mb}},\nonumber\\
&&\sigma(K^\pm p)|_{Pom.}=19.1 \, {\rm{mb}}.
\ear
%
%
This leads to
\bear
\label{41}
S_\pi&=& 0.60 \, {\rm{fm}},
\nonumber\\
S_K&=& 0.55 \, {\rm{fm}}.
\ear
%
%
%
\begin{figure}[htb]
 
\vspace*{-1.0cm}
 
\hspace{1cm}
\epsfysize=12cm
\centerline{\epsffile{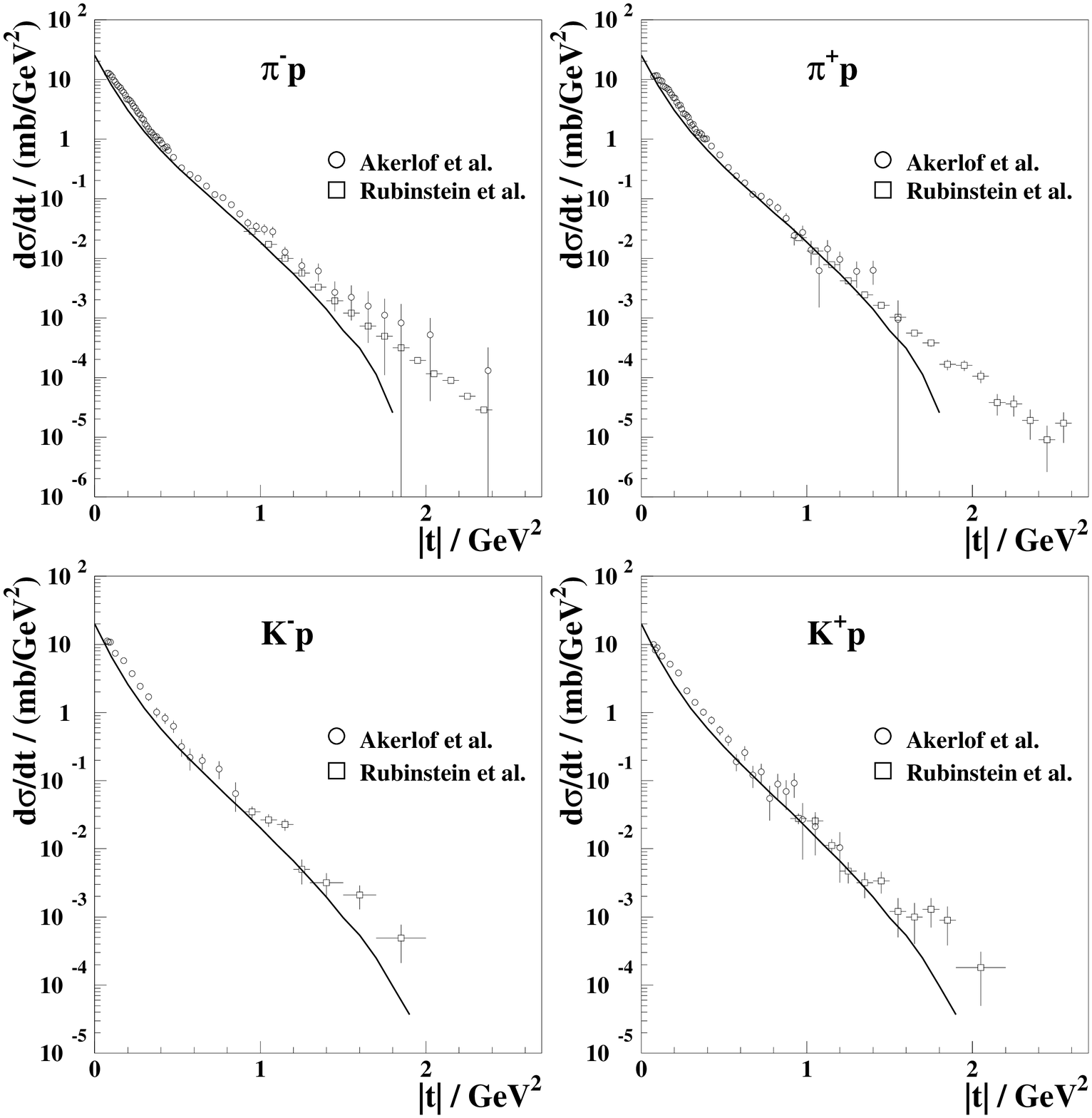}}
 
\vspace*{0cm}
 
\caption[a]{ \em Differential elastic cross sections for 
$ \pi^{\pm} p$ and $ K^{\pm} p$ scattering at 
$\sqrt{s}=19.5$ GeV.
This corresponds to 
$S_P=0.86$ {\rm{fm}}, $S_{\pi}=0.60$ {\rm{fm}} and $S_K=0.55$ 
{\rm{fm}}.
The data are from {\rm{\cite{mbdat}}}.  }
\label{mbbild}
\end{figure}
%
%
%
For comparison the electromagnetic radii are \cite{radii2}:
$r_{\pi}=0.66 \pm 0.01 \, {\rm fm},
r_{K}=0.58 \pm 0.04 \, {\rm fm}$.
Fig. \ref{mbbild} shows our results for the
$\pi^{\pm} p$ and $K^{\pm} p$
elastic differential cross sections.
As we can see the experimental data are again
reproduced quite well. As in $pp$ scattering the
slopes of the calculated cross sections are somewhat too large
for very small $|t|$. Furthermore all our $t$-distributions
are slightly below the experimental data up to $|t|\simeq 0.5\, GeV$.
This could again be due to with the missing real part
in the amplitude (\ref{mresult2}), a problem which might
be cured once we go beyond  the second
cumulant approximation.

We also note that the c.m. energy $\sqrt{s}=19.5\ GeV$
considered here is not very high and -- in Regge language --
effects of nonleading trajectories are still sizable and
more so for $\pi^-p,\ K^-p$
than for $\pi^+p,\ K^+p$ scattering. Our calculation does
not contain any non-leading Regge-exchanges and thus should
agree better with $\pi^+p,\ K^+p$ than $\pi^-p,\
K^-p$ scattering. This is not incompatible with the
results shown in Fig. \ref{mbbild}.

Due to the smaller extension parameters of the mesons our
calculated $t$-distributions for meson-proton scattering are
flatter than those for proton-proton scattering. Also
our calculation gives dips only around $|t|\approx 2 {\rm GeV}^2$
and we would not believe our model to be reliable at such
high $|t|$-values.

The squares of the ratios of our extension parameters
of mesons and the proton are
\bear\label{42}
 \left(\frac{S_\pi}{S_p}\right)^2=0.49, \;
 \left(\frac{S_K}{S_p}\right)^2=0.41, \;
 \left(\frac{S_K}{S_\pi}\right)^2=0.84.
\ear
The corresponding ratios for the mean squared  electromagnetic
radii are \cite{radii1,radii2}
\bear\label{43}
 \frac{r^2_\pi}{r^2_p}=0.55, \;
 \frac{r^2_K}{r^2_p}=0.42, \;
 \frac{r^2_K}{r^2_\pi}=0.77.
\ear
The ratios (\ref{42}) follow the trends of (\ref{43}) but certainly
are not equal to them. It is particularly noteworthy that in
our model the facts that (i) the $K^{\pm} p$ total cross sections
are smaller than the $\pi^\pm p$ ones and (ii) the $t$ distributions
for $K^\pm p$ flatter than for $\pi^\pm p$ are both reproduced
quantitatively by a smaller extension parameter $S_K$ compared
to $S_\pi$. In the additive quark model \cite{addrule}
on the other hand one has to assume a different cross section for the
scattering of $u,d$ on $u,d$ and $u,d$ on $s$ quarks which is hard
to understand since the gluon interaction is flavour-blind.

\section{Conclusion}

In this article we have presented calculations of amplitudes
for elastic
proton-proton, meson-meson, and meson-proton scattering at high
energies and small momentum transfer. Our model is based on
functional integral techniques \cite{na91} and an appropriate
evaluation of such integrals in the framework of the stochastic
vacuum model \cite{msv,dfk} In comparison with previous work
\cite{dfk,pir,15b} we have now made a cumulant expansion for the
correlation function of two Wegner-Wilson loops instead of an
expansion in terms of the number of field strength correlators.
We found that this latter expansion can only be justified for
medium and large impact parameters, but gives, nevertheless,
reasonable results for the total cross section and the slope
parameter at $t=0$.

As parameters in our model we have the QCD vacuum parameters: the
gluon condensate $G_2$, the non-abelian parameter $\kappa$, and
the correlation length $a$. In addition we have the $s$-dependent
hadron extension parameters $S_H(s)$.

We fixed the $S_H(s)$ by requiring the total cross sections
to agree with the experimental values. In this way we found quite
a good description of $d\sigma/dt$ for $pp,p\bar p,\pi^\pm p$ and
$K^\pm p$ scattering for vacuum parameters $G_2,\kappa, a$
or equivalently the string tension $\rho, \kappa$ and
$a$ as given in (\ref{31}-\ref{32a}). These values are close to the
corresponding values obtained from lattice data \cite{gitter, megg},
from previous investigations of high energy scattering \cite{dfk,mdoc}
and from various other low energy phenomena (see \cite{25a}
for a review). The parameters $S_H(s)$ came out close to the
known electromagnetic radii of the hadrons for energies $\sqrt s
\approx 20$ GeV and were required to increase slowly with $s$
according to (\ref{expar}).

As in previous work \cite{dfk} the fact that $\sigma_T(K p)$
is smaller than $\sigma_T(\pi p)$ is related in our model
to a smaller extension parameter $S_K$ compared to $S_\pi$.

The dip structure in $pp$ scattering around $|t|\approx 1.5$
GeV$^2$ was seen to depend crucially on the non-abelian character
of the gluon field strength correlator, i.e. on $\kappa\not=0$.
This leads to the string formation \cite{msv} and the
area law for the Wegner-Wilson loop with static quarks. In
our model these strings play again a crucial role in high
energy scattering in producing the correct $t$-distributions.

At very small values of $|t|$ our model gives a change of
slope related to the symmetric and antisymmetric combinations
of the multi-gluon exchange as explained after (\ref{mmatrix}).

As stated above one of the main ingredients of our model
is a cumulant expansion for the correlation function of 2 Wegner-Wilson
loops which we truncated after the second cumulant. But we can
easily see that the main features of our model will remain
unchanged if the matrix cumulant expansion 
(\ref{cumulantexpansion}) converges.
Then the complete sum of cumulants in (\ref{cumulantexpansion}) 
is a $9\times9$ matrix, invariant under $SU(3)$ rotations.
We will again obtain a decomposition for it as for
$C_2$ in (\ref{mchi2}) ff in terms of the projectors $P_s$
and $P_a$ with invariant functions $\chi_{s,a}$ multiplying
them etc. Of course the detailed shape of these functions
will be different.

Going back to our amplitude (\ref{mresult2}), we can
expand the cosines in powers of $\chi$. Maybe the $\chi^2$-term --
as kept in the work \cite{dfk} -- could be interpreted as
exchange of two nonperturbative gluons, as single bare
pomeron exchange, the $\chi^4$-term as exchange of
four nonperturbative gluons, as double pomeron exchange, etc.
At the moment, however, such an identification is purely
speculative.

In all this work we have treated baryons as quark-diquark
systems, i.e. we assumed 2 quarks of the 3 valence quarks
of a baryon to be close together in transverse directions. This
picture is supported e.g. by the investigation of \cite{odderon}
where it is shown to give an explanation for the apparent
absence of odderon couplings of the proton at small $|t|$. In
future work we plan to investigate meson-baryon and
baryon-baryon scattering treating baryons as 3-quark systems
with arbitrary distances between the quarks.
Also the dependence of the results
on the surfaces spanned into the loops (see Fig. \ref{pbild})
in order to apply the non-abelian Stokes theorem will be 
investigated.

To summarise: We have presented a model where the $t$-distributions
of elastic hadron-hadron scattering are related quantitatively
to the parameters of the vacuum in nonperturbative QCD.
The total cross sections which rise with energy were used
to fix the energy-dependent effective strong interaction extension
parameters of the hadrons. A theoretical calculation of
these extension parameters remains a challenge.

\section*{Acknowledgements}

We are grateful to H. G. Dosch, A. Donnachie, E. Ferreira, A.
Hebecker, W. Kilian, P. V. Landshoff, M. R\"uter, and Yu. A. Simonov
for many fruitful discussions. Special thanks are due to
E. Meggiolaro for clarifying discussions concerning the
lattice data and for extracting from the latter the vacuum
parameters in the form suitable for us.

\end{document}